\documentclass[11pt]{article}
\usepackage[dvips]{graphicx}
\usepackage{epsfig}
\usepackage{amssymb}
\usepackage{gensymb}
\usepackage{color}
\usepackage{amsmath}
\usepackage{units}      
\usepackage{cite}
\usepackage[left]{lineno}
\usepackage{color}

\newcommand{\kthreepic}[1]{K^{#1} \to \pi^{#1} \pi^{+} \pi^{-}}

\newcommand{\kefour}[1]{K^{#1} \to \pi^{+} \pi^{-} e^{#1} \nu}
\newcommand{\kmufour}[1]{K^{#1} \to \pi^{+} \pi^{-} \mu^{#1} \nu}
\newcommand{\kmunumumu}[1]{K^{#1} \to \mu^{+} \mu^{-} \mu^{#1} \nu}

\newcommand{\kmutwo}[1]{K^{#1} \to \mu^{#1} \nu}
\newcommand{\kmutwoN}[1]{K^{#1} \to \mu^{#1} N_{4}}
\newcommand{\kpichi}[1]{K^{#1} \to \pi^{#1} \chi}
\newcommand{\kpiX}[1]{K^{#1} \to \pi^{#1} X}
\newcommand{\Npimuws}{N_{4} \to \pi^{\mp}\mu^{\pm}}
\newcommand{\Npimurs}{N_{4} \to \pi^{\pm}\mu^{\mp}}
\newcommand{\Npimuns}{N_{4} \to \pi\mu}
\newcommand{\chimumu}{\chi \to \mu^{+}\mu^{-}}
\newcommand{\Xmumu}{X \to \mu^{+}\mu^{-}}

\newcommand{\kpimm}[1]{K^{#1} \to \pi^{#1} \mu^{+} \mu^{-}}
\newcommand{\kpimmws}{K^{\pm} \to \pi^{\mp} \mu^{\pm} \mu^{\pm}}
\newcommand{\kpimmns}[1]{K^{#1} \to \pi \mu \mu}

\newcommand{\kpmm}{K_{\pi\mu\mu}}
\newcommand{\kpmmlnv}{K_{\pi\mu\mu}^{\rm LNV}}
\newcommand{\kpmmlnc}{K_{\pi\mu\mu}^{\rm LNC}}
\newcommand{\npmmlnv}{N_{\pi\mu\mu}^{\rm LNV}}

\newcommand{\eqdef}{=}

\newcommand{\ttiny}[1]{\texttt{\tiny #1}}
\newcommand{\cc}{}
\def\cc/{\emph{c}}

\begin{document}
\centerline{\LARGE EUROPEAN ORGANIZATION FOR NUCLEAR RESEARCH}
%
%
\vspace{15mm}
{\flushright{
CERN-EP-2016-310 \\
14 March 2017\\
}}
\vspace{40mm}
%
\vspace{10mm}

\begin{center}
\boldmath
{\bf {\Large \boldmath{Searches for Lepton Number Violation and Resonances\\in $K^{\pm}\to\pi\mu\mu$ Decays}}}
\unboldmath
\end{center}
\begin{center}
{\Large The NA48/2 collaboration}\\
\end{center}

\begin{abstract}
The NA48/2 experiment at CERN collected a large sample of charged kaon decays to final states with multiple charged particles in 2003--2004.
A new upper limit on the rate of the lepton number violating decay $K^{\pm}\to\pi^{\mp}\mu^{\pm}\mu^{\pm}$ is reported: $\mathcal{B}(K^{\pm}\to\pi^{\mp}\mu^{\pm}\mu^{\pm})<8.6 \times 10^{-11}$ at 90\% CL. Searches for two-body resonances~$X$ in $K^{\pm}\to\pi\mu\mu$ decays (such as heavy neutral leptons~$N_4$ and inflatons~$\chi$) are also presented. 
In the absence of signals, upper limits are set on the products of branching fractions~$\mathcal{B}(K^{\pm}\to\mu^{\pm}N_4)\mathcal{B}(N_4\to\pi\mu)$ and $\mathcal{B}(K^{\pm}\to\pi^{\pm}X)\mathcal{B}(X\to\mu^+\mu^-)$ for ranges of assumed resonance masses and lifetimes. The limits are in the $(10^{-11},10^{-9})$ range for resonance lifetimes below~$\unit[100]{ps}$.
\end{abstract}

\begin{center}
{\it Accepted for publication in Physics Letters B}
\end{center}

\newpage
\setlength{\topmargin}{-18mm}
\begin{center}
{\Large The NA48/2 Collaboration}\\
\vspace{2mm}
 J.R.~Batley,
 G.~Kalmus,
 C.~Lazzeroni$\,$\footnotemark[1]$^,$\footnotemark[2],
 D.J.~Munday,
 M.W.~Slater$\,$\footnotemark[1],
 S.A.~Wotton \\
{\em \small Cavendish Laboratory, University of Cambridge,
Cambridge, CB3 0HE, UK$\,$\footnotemark[3]} \\[0.2cm]
 R.~Arcidiacono$\,$\footnotemark[4],
 G.~Bocquet,
 N.~Cabibbo$\,$\renewcommand{\thefootnote}{\fnsymbol{footnote}}%
\footnotemark[2]\renewcommand{\thefootnote}{\arabic{footnote}},
 A.~Ceccucci,
 D.~Cundy$\,$\footnotemark[5],
 V.~Falaleev,
 M.~Fidecaro,
 L.~Gatignon,
 A.~Gonidec,
 W.~Kubischta,
 A.~Maier,
 K.~Massri$\,$\renewcommand{\thefootnote}{\fnsymbol{footnote}}%
\footnotemark[1]\renewcommand{\thefootnote}{\arabic{footnote}},\\
 A.~Norton$\,$\footnotemark[6],
 M.~Patel$\,$\footnotemark[7],
 A.~Peters\\
{\em \small CERN, CH-1211 Gen\`eve 23, Switzerland} \\[0.2cm]
 S.~Balev$\,$\renewcommand{\thefootnote}{\fnsymbol{footnote}}%
\footnotemark[2]\renewcommand{\thefootnote}{\arabic{footnote}},
 P.L.~Frabetti,
 E.~Gersabeck$\,$\footnotemark[8],
 E.~Goudzovski$\,$\renewcommand{\thefootnote}{\arabic{footnote}}\footnotemark[1]$^,$\footnotemark[2]$^,$\footnotemark[9],
 P.~Hristov$\,$\footnotemark[10],
 V.~Kekelidze,
 V.~Kozhuharov$\,$\footnotemark[11]$^,$\footnotemark[12],
 L.~Litov$\,$\footnotemark[11],
 D.~Madigozhin,
 N.~Molokanova,
 I.~Polenkevich,
 Yu.~Potrebenikov,
 S.~Stoynev$\,$\footnotemark[13],
 A.~Zinchenko$\,$\renewcommand{\thefootnote}{\fnsymbol{footnote}}%
\footnotemark[2]\renewcommand{\thefootnote}{\arabic{footnote}} \\
{\em \small Joint Institute for Nuclear Research, 141980 Dubna (MO), Russia} \\[0.2cm]
 E.~Monnier$\,$\footnotemark[14],
 E.~Swallow$\,$\renewcommand{\thefootnote}{\fnsymbol{footnote}}%
\footnotemark[2]\renewcommand{\thefootnote}{\arabic{footnote}},
 R.~Winston$\,$\footnotemark[15]\\
{\em \small The Enrico Fermi Institute, The University of Chicago,
Chicago, IL 60126, USA}\\[0.2cm]
 P.~Rubin$\,$\footnotemark[16],
 A.~Walker \\
{\em \small Department of Physics and Astronomy, University of
Edinburgh, Edinburgh, EH9 3JZ, UK} \\[0.2cm]
 W.~Baldini,
 A.~Cotta Ramusino,
 P.~Dalpiaz,
 C.~Damiani,
 M.~Fiorini,
 A.~Gianoli, \\
 M.~Martini,
 F.~Petrucci,
 M.~Savri\'e,
 M.~Scarpa,
 H.~Wahl \\
 {\em \small Dipartimento di Fisica e Scienze della Terra dell'Universit\`a e Sezione
dell'INFN di Ferrara, \\ I-44122 Ferrara, Italy} \\[0.2cm]
 A.~Bizzeti$\,$\footnotemark[17],
 M.~Lenti,
 M.~Veltri$\,$\footnotemark[18] \\
{\em \small Sezione dell'INFN di Firenze, I-50019 Sesto Fiorentino, Italy} \\[0.2cm]
 M.~Calvetti,
 E.~Celeghini,
 E.~Iacopini,
 G.~Ruggiero$\,$\footnotemark[19] \\
{\em \small Dipartimento di Fisica dell'Universit\`a e Sezione
dell'INFN di Firenze, I-50019 Sesto Fiorentino, Italy} \\[0.2cm]
 M.~Behler,
 K.~Eppard,
 K.~Kleinknecht,
 P.~Marouelli,
 L.~Masetti,
 U.~Moosbrugger,\\
 C.~Morales Morales$\,$\footnotemark[20],
 B.~Renk,
 M.~Wache,
 R.~Wanke,
 A.~Winhart$\,$\footnotemark[1]\\
{\em \small Institut f\"ur Physik, Universit\"at Mainz, D-55099 Mainz, Germany$\,$\footnotemark[21]} \\[0.2cm]
 D.~Coward$\,$\footnotemark[22],
 A.~Dabrowski$\,$\footnotemark[10],
 T.~Fonseca Martin,
 M.~Shieh,
 M.~Szleper$\,$\footnotemark[23],\\
 M.~Velasco,
 M.D.~Wood$\,$\footnotemark[22] \\
{\em \small Department of Physics and Astronomy, Northwestern
University, Evanston, IL 60208, USA}\\[0.2cm]
 P.~Cenci,
 M.~Pepe,
 M.C.~Petrucci \\
{\em \small Sezione dell'INFN di Perugia, I-06100 Perugia, Italy} \\[0.2cm]
 G.~Anzivino,
 E.~Imbergamo,
 A.~Nappi$\,$\renewcommand{\thefootnote}{\fnsymbol{footnote}}%
\footnotemark[2]\renewcommand{\thefootnote}{\arabic{footnote}},
 M.~Piccini,
 M.~Raggi$\,$\footnotemark[24],
 M.~Valdata-Nappi \\
{\em \small Dipartimento di Fisica dell'Universit\`a e
Sezione dell'INFN di Perugia, I-06100 Perugia, Italy} \\[0.2cm]
 C.~Cerri,
 R.~Fantechi \\
{\em Sezione dell'INFN di Pisa, I-56100 Pisa, Italy} \\[0.2cm]
 G.~Collazuol$\,$\footnotemark[25],
 L.~DiLella$\,$\footnotemark[26],
 G.~Lamanna$\,$\footnotemark[26],
 I.~Mannelli,
 A.~Michetti \\
{\em Scuola Normale Superiore e Sezione dell'INFN di Pisa, I-56100
Pisa, Italy} \\[0.2cm]
 F.~Costantini,
 N.~Doble,
 L.~Fiorini$\,$\footnotemark[27],
 S.~Giudici,
 G.~Pierazzini$\,$\renewcommand{\thefootnote}{\fnsymbol{footnote}}%
\footnotemark[2]\renewcommand{\thefootnote}{\arabic{footnote}},\\
 M.~Sozzi, S.~Venditti\\
{\em Dipartimento di Fisica dell'Universit\`a e Sezione dell'INFN di
Pisa, I-56100 Pisa, Italy} \\[0.2cm]
\newpage
 B.~Bloch-Devaux$\,$\footnotemark[28],
 C.~Cheshkov$\,$\footnotemark[29],
 J.B.~Ch\`eze,
 M.~De Beer,
 J.~Derr\'e,
 G.~Marel,
 E.~Mazzucato,
 B.~Peyaud,
 B.~Vallage \\
{\em \small DSM/IRFU -- CEA Saclay, F-91191 Gif-sur-Yvette, France} \\[0.2cm]
 M.~Holder,
 M.~Ziolkowski \\
{\em \small Fachbereich Physik, Universit\"at Siegen, D-57068 Siegen, Germany$\,$\footnotemark[30]} \\[0.2cm]
 C.~Biino,
 N.~Cartiglia,
 F.~Marchetto \\
{\em \small Sezione dell'INFN di Torino, I-10125 Torino, Italy} \\[0.2cm]
 S.~Bifani$\,$\footnotemark[1],
 M.~Clemencic$\,$\footnotemark[10],
 S.~Goy Lopez$\,$\footnotemark[31]\\
{\em \small Dipartimento di Fisica dell'Universit\`a e
Sezione dell'INFN di Torino, I-10125 Torino, Italy} \\[0.2cm]
 H.~Dibon,
 M.~Jeitler,
 M.~Markytan,
 I.~Mikulec,
 G.~Neuhofer,
 L.~Widhalm$\,$\renewcommand{\thefootnote}{\fnsymbol{footnote}}%
\footnotemark[2]\renewcommand{\thefootnote}{\arabic{footnote}} \\
{\em \small \"Osterreichische Akademie der Wissenschaften, Institut
f\"ur Hochenergiephysik,\\ A-10560 Wien, Austria$\,$\footnotemark[32]} \\[0.5cm]
\end{center}

%
\renewcommand{\thefootnote}{\fnsymbol{footnote}}
\footnotetext[1]{Corresponding author, email: karim.massri@cern.ch}
\footnotetext[2]{Deceased}
\renewcommand{\thefootnote}{\arabic{footnote}}
\footnotetext[1]{Now at: School of Physics and Astronomy, University of Birmingham, Birmingham, B15 2TT, UK}
\footnotetext[2]{Supported by a Royal Society University Research Fellowship (UF100308, UF0758946)}
\footnotetext[3]{Funded by the UK Particle Physics and Astronomy Research Council, grant PPA/G/O/1999/00559}
\footnotetext[4]{Now at: Universit\`a degli Studi del Piemonte Orientale e Sezione dell'INFN di Torino, I-10125 Torino, Italy}
\footnotetext[5]{Now at: Istituto di Cosmogeofisica del CNR di Torino, I-10133 Torino, Italy}
\footnotetext[6]{Now at: Dipartimento di Fisica e Scienze della Terra dell'Universit\`a e Sezione dell'INFN di Ferrara, I-44122 Ferrara, Italy}
\footnotetext[7]{Now at: Department of Physics, Imperial College, London, SW7 2BW, UK}
\footnotetext[8]{Now at: Physikalisches Institut, Ruprecht-Karls-Universit\"at Heidelberg, D-69120 Heidelberg, Germany}
\footnotetext[9]{Supported by ERC Starting Grant 336581}
\footnotetext[10]{Now at: CERN, CH-1211 Gen\`eve 23, Switzerland}
\footnotetext[11]{Now at: Faculty of Physics, University of Sofia ``St. Kl. Ohridski'', 1164 Sofia, Bulgaria, funded by the Bulgarian National Science Fund under contract DID02-22}
\footnotetext[12]{Also at Laboratori Nazionali di Frascati, I-00044 Frascati, Italy}
\footnotetext[13]{Now at: Fermi National Accelerator Laboratory, Batavia, IL 60510, USA}
\footnotetext[14]{Now at: Centre de Physique des Particules de Marseille, IN2P3-CNRS, Universit\'e de la M\'editerran\'ee, F-13288 Marseille, France}
\footnotetext[15]{Now at: School of Natural Sciences, University of California, Merced, CA 95344, USA}
\footnotetext[16]{Now at: School of Physics, Astronomy and Computational Sciences, George Mason
University, Fairfax, VA 22030, USA}
\footnotetext[17]{Also at Dipartimento di Scienze Fisiche, Informatiche e Matematiche, Universit\`a di Modena e Reggio Emilia, I-41125 Modena, Italy}
\footnotetext[18]{Also at Istituto di Fisica, Universit\`a di Urbino, I-61029 Urbino, Italy}
\footnotetext[19]{Now at: Department of Physics, University of Liverpool, Liverpool, L69 7ZE, UK}
\footnotetext[20]{Now at: Helmholtz-Institut Mainz, Universit\"at Mainz, D-55099 Mainz, Germany}
\footnotetext[21]{Funded by the German Federal Minister for Education and Research under contract 05HK1UM1/1}
\footnotetext[22]{Now at: SLAC, Stanford University, Menlo Park, CA 94025, USA}
\footnotetext[23]{Now at: National Centre for Nuclear Research, P-05-400 \'Swierk, Poland}
\footnotetext[24]{Now at: Universit\`a di Roma ``La Sapienza", I-00185 Roma, Italy}
\footnotetext[25]{Now at: Dipartimento di Fisica dell'Universit\`a e Sezione dell'INFN di Padova, I-35131 Padova, Italy}
\footnotetext[26]{Now at: Dipartimento di Fisica dell'Universit\`a e Sezione dell'INFN di Pisa, I-56100 Pisa, Italy}
\footnotetext[27]{Now at: Instituto de F\'isica Corpuscular IFIC, Universitat de Valencia, E-46071 Valencia, Spain}
\footnotetext[28]{Now at: Dipartimento di Fisica dell'Universit\`a di Torino, I-10125 Torino, Italy}
\footnotetext[29]{Now at: Institut de Physique Nucl\'eaire de Lyon, Universit\'e Lyon I, F-69622 Villeurbanne, France}
\footnotetext[30]{Funded by the German Federal Minister for Research and Technology (BMBF) under contract 056SI74}
\footnotetext[31]{Now at: Centro de Investigaciones Energeticas
Medioambientales y Tecnologicas, E-28040 Madrid, Spain}
\footnotetext[32]{Funded by the Austrian Ministry for Traffic and Research under the contract GZ 616.360/2-IV GZ 616.363/2-VIII, and by the Fonds f\"ur Wissenschaft und Forschung FWF Nr.~P08929-PHY}

\newpage

\boldmath
\section*{Introduction}
\unboldmath
Neutrinos are strictly massless within the Standard Model~(SM), due to the absence of right\mbox{-}hand\-ed neutrino states. However, since the observation of neutrino oscillations has unambiguously demonstrated the massive nature of neutrinos, right-handed neutrino states must be included.
A natural extension of the SM involves the inclusion of sterile neutrinos which mix with ordinary neutrinos: an example is the Neutrino Minimal Standard Model~($\nu$MSM)~\cite{as05_01,as05_02}.
In this model, three massive right-handed neutrinos are introduced to explain neutrino oscillations, dark matter and baryon asymmetry of the Universe: the lightest one with a mass~$\mathcal{O}(1\mbox{ keV}/c^2)$ is a dark matter candidate; the other two with masses~$\mathcal{O}(100\mbox{ MeV}/c^2)$ are responsible for the masses of the SM neutrinos (via the see-saw mechanism) and introduce extra CP violating phases to account for baryon asymmetry.
The $\nu$MSM can be further extended by adding a scalar field to incorporate inflation and provide a common source for electroweak symmetry breaking and right-handed neutrino masses~\cite{sh06}.
The new particles predicted by these models can be produced in kaon decays.
In particular, the Lepton Number Viola\-ting~(LNV) $\kpimmws$ decay forbidden in the SM 
could proceed via an off-shell or an on-shell Majorana neutrino~$N_4$~\cite{li00,at09}, while an inflaton~$\chi$ could be produced in the Lepton Number Conserving~(LNC) $K^{\pm}\to\pi^{\pm}\chi$ decay, and decay promptly to $\chi\to\mu^+\mu^-$~\cite{be10,be14}.

The currently most stringent contraint on the branching fraction $\mathcal{B}(\kpimmws)$ has been established by the NA48/2 experiment~\cite{ba11}, improving on the previous limit set by the BNL-E865 experiment~\cite{ap00}.
Limits on the heavy neutrino coupling~$|U_{\mu4}|$ from neutrino decay searches have been obtained by beam dump~\cite{co85,ba86,be88,ad92,ga95,vi95,ab97,va99} and $B$ decay~\cite{aa14,sh16} experiments, while the constraints on the inflaton mixing angle~$\theta$ have been set by a phenomenological study of beam dump and $B$ decay experimental results~\cite{be13}. A stringent constraint from a dedicated search for inflatons in $B$ decays has been published recently~\cite{aa16}.

This letter reports a search for the LNV $\kpimmws$ decay and two-body resonances in $\kpimmns{\pm}$ decays using a sample of $K^\pm$ decays
collected by the NA48/2 experiment at CERN in 2003--2004. The experiment was exposed to about $2\times 10^{11}$ $K^\pm$ decays.
The substantial improvement in the search for the $\kpimmws$ decay with respect to the analysis reported in Ref.~\cite{ba11} is due to the use of an event selection developed specifically for background suppression and a muon reconstruction optimized to increase the acceptance for events with multiple muons, which was not required to obtain the main result of Ref.~\cite{ba11}.

\boldmath
\section{Beam, detector and data sample}
\unboldmath
\label{sec:experiment}

The NA48/2 experiment used simultaneous $K^+$ and $K^-$ beams produced by 400~GeV/$c$ primary CERN SPS protons impinging on a beryllium target. Charged particles with momenta of ${(60\pm3)}$~GeV/$c$ were selected by an achromatic system of four dipole magnets which split the two beams in the vertical plane and recombined them on a common axis. The beams then passed through collimators and a series of quadrupole magnets, and entered a 114~m long cylindrical vacuum tank with a diameter of 1.92~m to 2.4~m containing the fiducial decay region. Both beams had an angular divergence of about 0.05~mrad, a transverse size of about 1~cm, and were aligned with the longitudinal axis of the detector within 1~mm.

The vacuum tank was followed by a magnetic spectrometer housed in a vessel filled with helium at nearly atmospheric pressure, separated from the vacuum by a thin ($0.3\%~X_0$) $\rm{Kevlar}\textsuperscript{\textregistered}$ window.
An aluminium beam pipe of 158~mm outer diameter traversing the centre of the spectrometer (and all the following detectors) allowed the undecayed beam particles to continue their path in vacuum. The spectrometer consisted of four drift chambers (DCH) with a transverse size of 2.9~m: DCH1, DCH2 located upstream and DCH3, DCH4 downstream of a dipole magnet that provided a horizontal transverse momentum kick of 120~MeV/$c$ for charged particles.
Each DCH was composed of four staggered double planes of sense wires to measure $X(0\degree)$, $Y(90\degree)$, $U$ and $V(\pm 45\degree)$ coordinates.
The DCH space point reso\-lution was 90~$\mu$m in both horizontal and vertical directions, and the momentum resolution was $\sigma_p/p = (1.02 \oplus 0.044\cdot p)\%$, with $p$ expressed in GeV/$c$. The spectrometer was followed by a plastic scintillator hodoscope (HOD) with a transverse size of about 2.4 m, consisting of a plane of vertical and a plane of horizontal strip-shaped counters arranged in four quadrants (each divided logically into four regions). The HOD provided time measurements for charged particles with 150~ps resolution. It was followed by a liquid krypton electromagnetic calorimeter~(LKr), an almost homogeneous ioni\-zation chamber with an active volume of 7 m$^3$, $27~X_0$ deep, segmented transversally into 13248 projective $\sim\!2\!\times\!2$~cm$^2$ cells. The LKr energy resolution was $\sigma_E/E=(3.2/\sqrt{E}\oplus9/E\oplus0.42)\%$, the spatial resolution for an isolated electromagnetic shower was $(4.2/\sqrt{E}\oplus0.6)$~mm in both horizontal and vertical directions, and the time resolution was $2.5~{\rm ns}/\sqrt{E}$, with $E$ expressed in GeV. 
The LKr was followed by a hadronic calorimeter, which was an iron-scintillator sandwich with a total iron thickness of 1.2~m. A muon detector (MUV), located further downstream, consisted of three $2.7\times2.7$~m$^2$ planes of plastic scintillator strips, each preceded by a 80~cm thick iron wall. 
The strips (aligned horizontally in the first and last planes, vertically in the middle plane) were 2.7~m long and 2~cm thick, and read out by photomultipliers at both ends. The first two planes contained 11 strips, while the third plane consisted of 6 strips.
A detailed description of the beamline and the detector can be found in Refs.~\cite{fa07,ba07}.

The NA48/2 experiment collected data in 2003--2004, with about 100 days of effective data taking in total. A two-level trigger chain was employed to collect $K^\pm$ decays with at least three charged tracks in the final state, originating from the same vertex. At the first level~(L1), a coincidence of hits in the two planes of the HOD was required in at least two of the 16 non\mbox{-}overlapping logical regions. The second level (L2) performed online reconstruction of trajectories and momenta of charged particles based on the DCH information. The L2 logic was based on the multiplicities and kinematics of reconstructed tracks and two-track vertices. The overall trigger efficiency for three-track kaon decays was above 98.5\%~\cite{ba07}.

A GEANT3-based~\cite{geant} Monte Carlo (MC) simulation including full beamline, detector geo\-metry and material description, magnetic fields, local inefficiencies, misalignment and their time variations throughout the running period is used to evaluate the detector response.


\boldmath
\section{Event reconstruction and selection}
\unboldmath
\label{sec:selection}
Three-track vertices (compatible with either $\kpimmns{\pm}$ or $K^\pm\to\pi^\pm\pi^+\pi^-$ decay topology, denoted $\kpmm$ and $K_{3\pi}$ below) are reconstructed by extrapolation of track segments from the spectrometer upstream into the decay region, taking into account the measured Earth's magnetic field, stray fields due to magnetization of the vacuum tank, and multiple scattering. Within the 50~cm resolution on the longitudinal vertex position, $\kpimmws$ and $\kpimm{\pm}$ decays (denoted $\kpmmlnv$ and $\kpmmlnc$ below) mediated by short-lived (lifetime $\tau\lesssim \unit[10]{ps}$) particles are indistinguishable from three-track decays.

The $\kpmm$ decay rates are measured relative to the abundant $K_{3\pi}$ normalization channel. The $\kpmm$ and $K_{3\pi}$ samples have been collected concurrently using the same trigger logic. The fact that the $\mu^\pm$ and $\pi^\pm$ masses are close results in similar topologies of the signal and normalization final states. This leads to first order cancellation of the systematic effects induced by imperfect kaon beam description, local detector inefficiencies, and trigger inefficiency.
The selection procedures for the $\kpmm$ and $K_{3\pi}$ modes have a large common part, namely the requirement of a reconstructed three-track vertex satisfying the following main~criteria.
\begin{itemize}
\item The total charge of the three tracks is $Q=\pm1$.
\item The vertex is located within the 98~m long fiducial decay region, which starts 2~m downstream of the beginning of the vacuum tank.
\item The vertex track momenta~$p_i$ are within the range $(5,55)~{\rm GeV}/c$, and the total momentum of the three tracks $|\sum\vec p_i|$ is consistent with the beam nominal range of $(55,65)~{\rm GeV}/c$.
\item The total transverse momentum of the three tracks with respect to the actual beam direction (which is measured with the $K_{3\pi}$ sample) is $p_T<10~{\rm MeV}/c$.
\end{itemize}
If several vertices satisfy the above conditions, the one with the lowest fit $\chi^2$ is considered. 
The tracks forming the vertex are required to satisfy the following conditions.
\begin{itemize}
 \item Tracks are consistent in time (within 10~ns from the average time of the three tracks) and with the trigger time.
 \item Tracks are in the DCH, HOD, LKr and MUV geometric acceptances.
 \item Track separations exceed 2~cm in the DCH1 plane to suppress photon conversions, and 20~cm in the LKr and MUV front planes to minimize particle misidentification due to shower overlaps and multiple scattering.
\end{itemize}
The $\kpmmlnv$ ($\kpmmlnc$) candidates are then selected using the particle identification and kinematic criteria listed below.
\begin{itemize}
\item The vertex is required to be composed of one $\pi^\pm$ candidate, with the ratio of energy deposition in the LKr calorimeter to momentum measured by the spectrometer ${E/p<0.95}$ to suppress electrons ($e^{\pm}$) and no in-time associated hits in the MUV, and a pair of identically (oppositely) charged $\mu^\pm$ candidates, with $E/p<0.2$ and associated hits in the first two planes of the MUV. 
The $\pi^{\pm}$ candidate is required to have momentum above $15~{\rm GeV}/c$ to ensure a high muon suppression factor, measured from reconstructed ${\kmutwo{\pm}}$~decays to increase with momentum and to be 40~(125) at $p=10$~(15)~GeV/$c$.
\item The invariant mass of the three tracks in the $\pi^\mp\mu^\pm\mu^\pm$ ($\pi^\pm\mu^+\mu^-$) hypothesis satisfies
$|M_{\pi\mu\mu}-M_K|< \mbox{5 (8) MeV}/c^2$, where $M_K$ is the nominal $K^{\pm}$ mass~\cite{pdg}. This interval corresponds to $\pm 2$ ($\pm3.2$) times the resolution $\sigma_{\pi\mu\mu} = 2.5$~MeV/$c^2$.
The different signal region definition between $\kpmmlnv$ and $\kpmmlnc$ selections is a result of the optimization of the expected sensitivities, due to the different background composition (Sec.~\ref{sec:datasamples}).
\item When searching for resonances: $|M_{ij}-M_{X}|<\delta_M(M_{X})$, where $M_{ij}$ is the invariant mass of the $ij$ pair ($ij = \pi\mu, \mu^+\mu^-$), $M_X$ is the assumed resonance mass, and the \mbox{half-width}~$\delta_M(M_{X})$ of the resonance search window, depending on $M_{X}$, is defined in Sec.~\ref{sec:res_search}.
Two possible values for $M_{\pi\mu}$ exist in the $\kpmmlnv$ selection, since the muon produced by the $K^{\pm}$ decay cannot be distinguished from the one produced by the subsequent $N_4$ decay. In this case, the value that minimizes $|M_{\pi\mu}-M_X|$ is considered.
\end{itemize}
Independently, the following criteria are applied to select the $K_{3\pi}$ decays.
\begin{itemize}
\item The pion identification criteria described above are applied only to the track with the electric charge opposite to that of the kaon, to symmetrize the selection of the signal and normalization modes and diminish the corresponding systematic uncertainties.
\item The invariant mass of the three tracks in the $3\pi^\pm$ hypothesis satisfies ${|M_{3\pi}-M_K|} <5~{\rm MeV}/c^2$, which corresponds approximately to $\pm 3$ times the resolution $\sigma_{3\pi} = 1.7$~MeV/$c^2$.
\end{itemize}
No restrictions are applied on additional energy deposition in the LKr calorimeter and extra tracks not belonging to the vertex, to decrease the sensitivity to accidental activity.
To avoid bias during the choice of the event selection criteria, the $\kpmmlnv$ selection was optimized with a blind analysis: 
an independent $K_{3\pi}$ MC sample was used to study the $K_{3\pi}$ background suppression; 
furthermore, the data events with invariant mass~$M_{\pi\mu\mu}$ satisfying $|M_{\pi\mu\mu}-M_K|< 10~{\rm MeV}/c^2$ were discarded and the data/MC agreement was studied in the $M_{\pi\mu\mu}$ control region $456\mbox{ MeV}/c^2<M_{\pi\mu\mu}<480\mbox{ MeV}/c^2$.

\boldmath
\section{Data and MC samples}
\unboldmath
\label{sec:datasamples}
The number of $K^\pm$ decays in the 98~m long fiducial decay region is measured as
$$
N_K = \frac{N_{3\pi}\cdot D}{{\cal B}(K_{3\pi})A(K_{3\pi})} = (1.637\pm0.007)\times 10^{11},
$$
where $N_{3\pi} = 1.367\times10^{7}$ is the number of $K_{3\pi}$ candidates reconstructed in the data sample (with a negligible background contamination), $D = 100$ is the downscaling factor of the $K_{3\pi}$ subset used for the $N_K$ measurement, ${\cal B}(K_{3\pi})$ is the nominal branching fraction of the $K_{3\pi}$ decay mode~\cite{pdg} and $A(K_{3\pi}) = 14.96\%$ is the acceptance of the selection evaluated with MC simulations.
The main contribution to the quoted uncertainty of $N_K$ is due to the external error on~${\cal B}(K_{3\pi})$.

\begin{figure}[t!]
\begin{center}
\resizebox{0.5\textwidth}{!}{\includegraphics{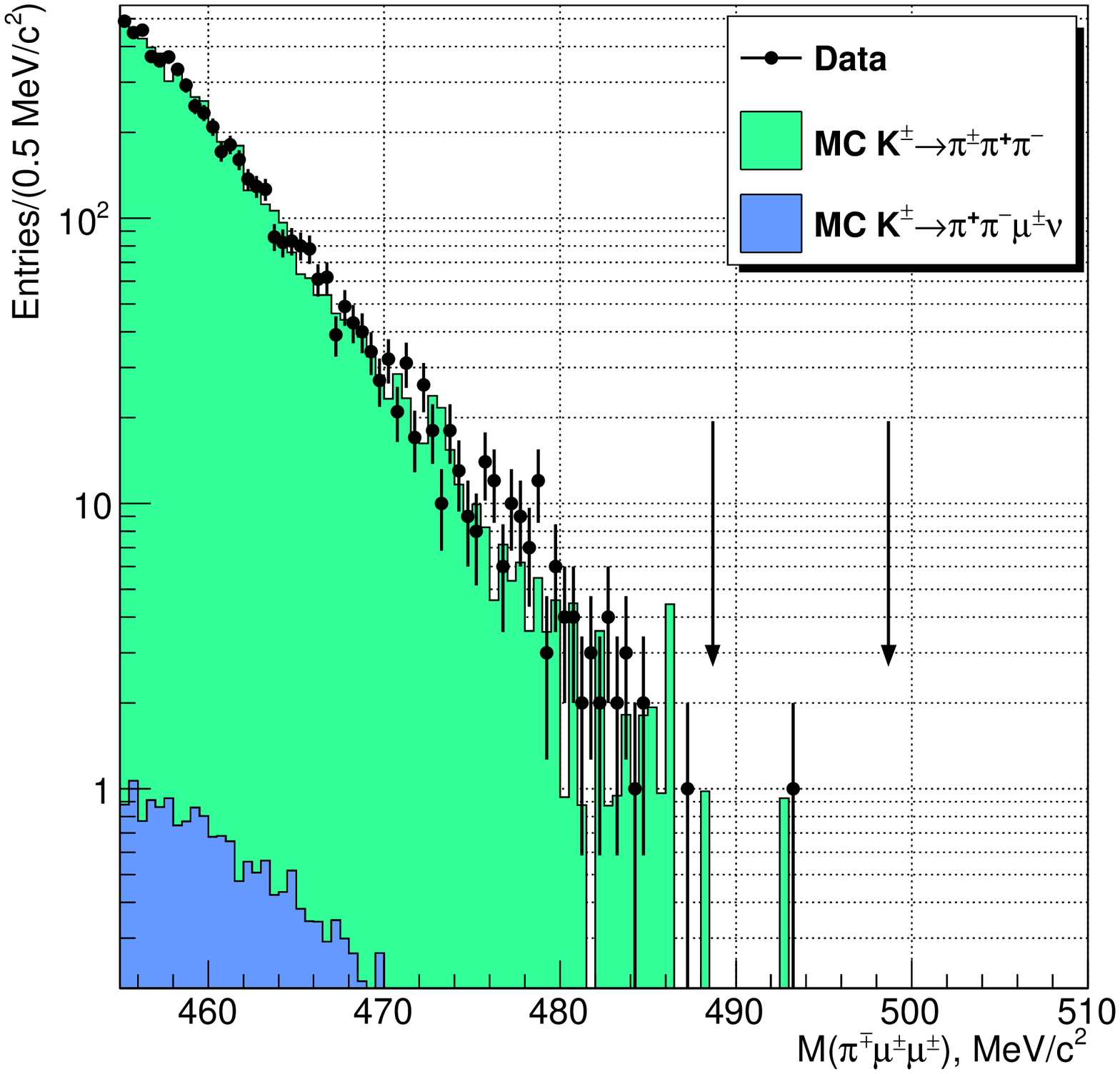}}%
\put(-25,203){(a)}
\resizebox{0.5\textwidth}{!}{\includegraphics{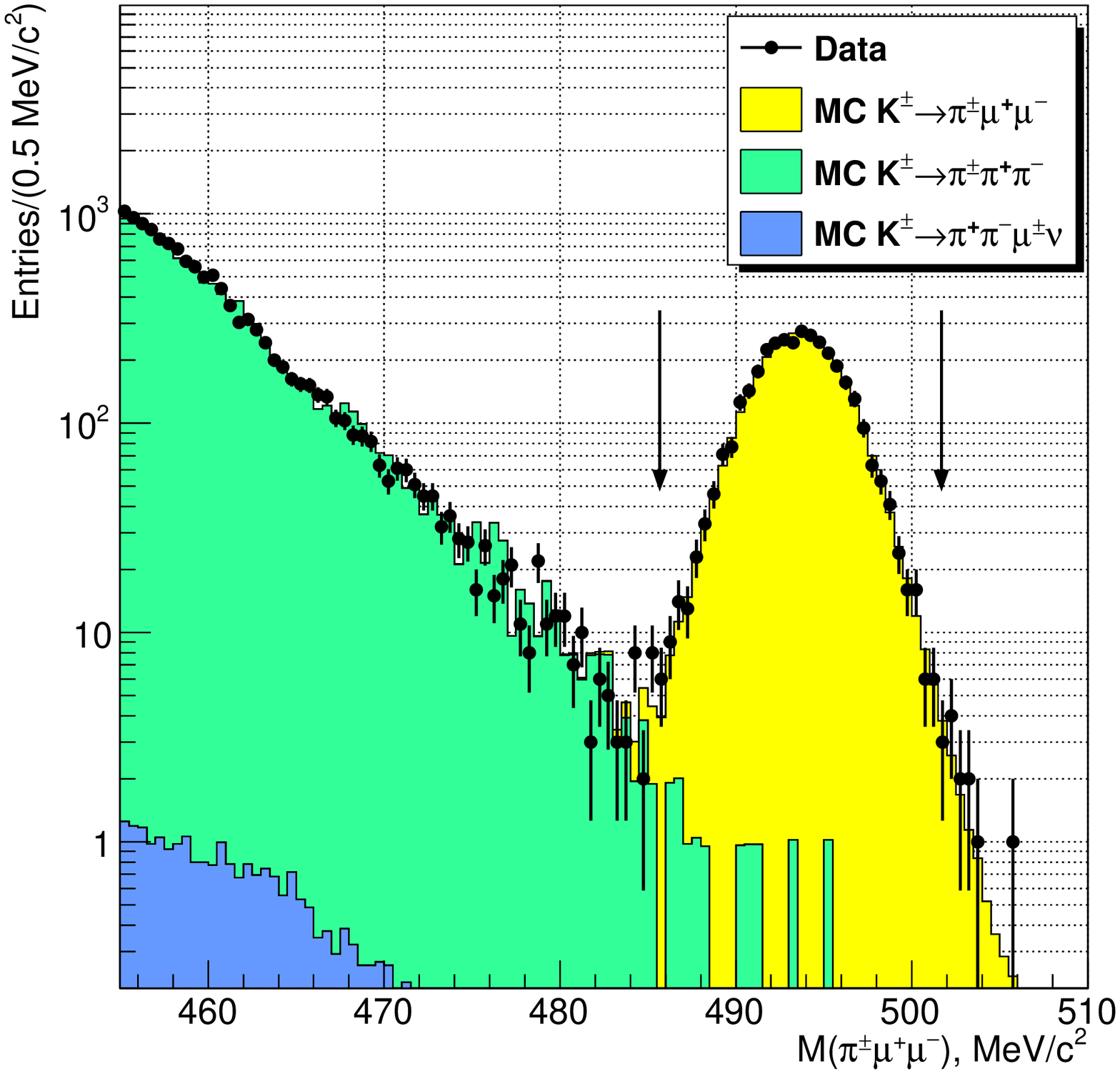}}
\put(-25,203){(b)}
\\
\hspace{0.01\textwidth}\resizebox{0.49\textwidth}{!}{\includegraphics{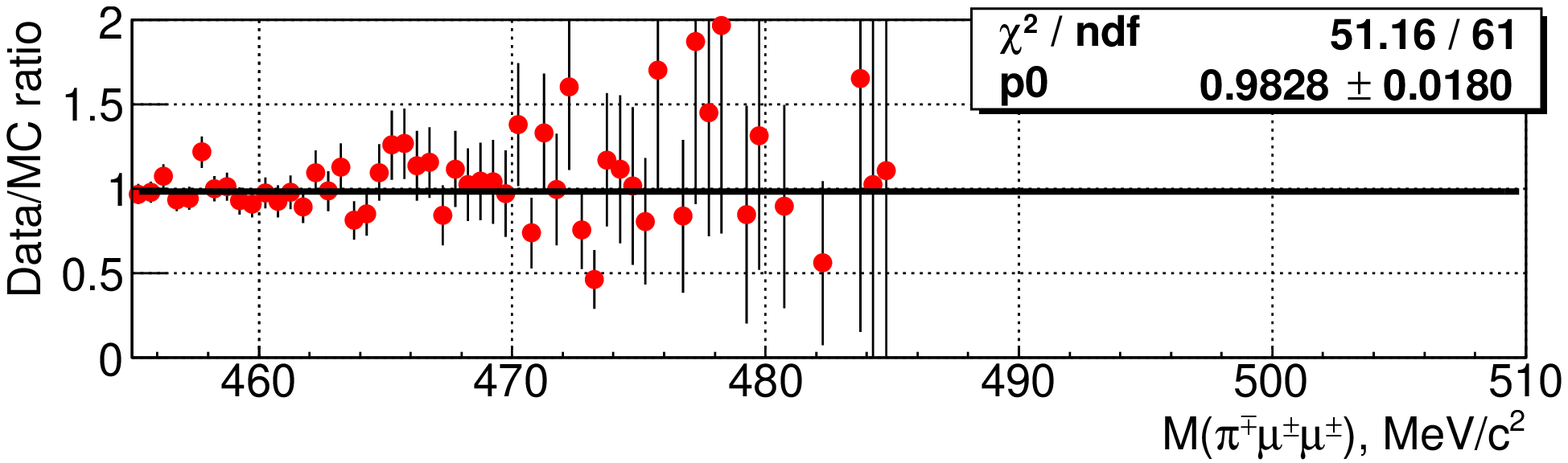}}%
\hspace{0.01\textwidth}\resizebox{0.49\textwidth}{!}{\includegraphics{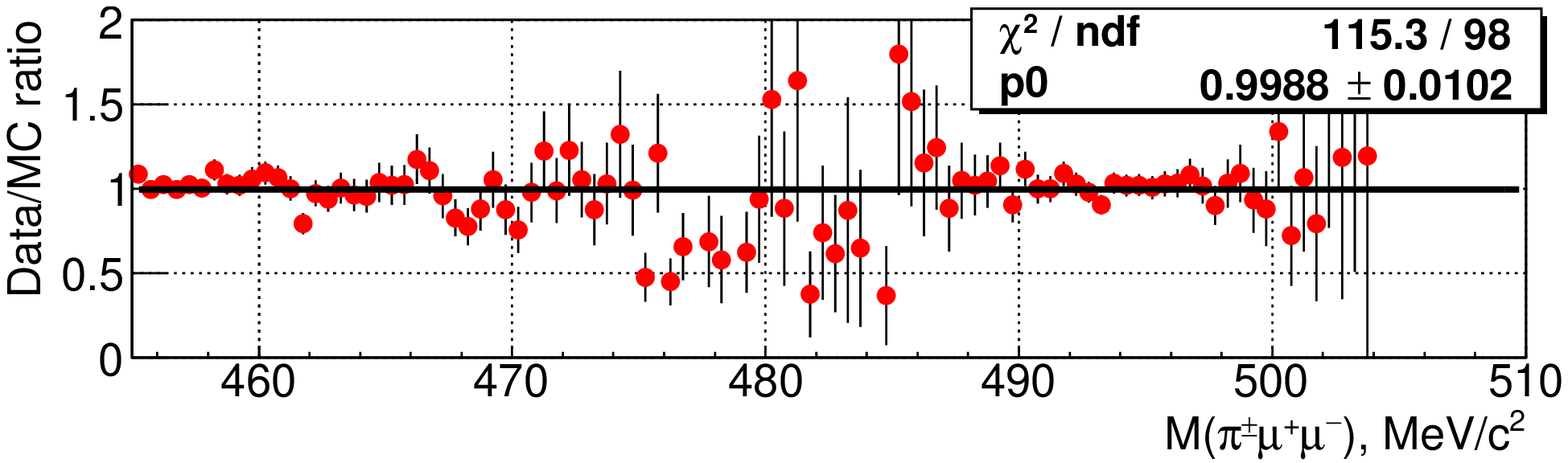}}%
\end{center}
\vspace{-13mm}
\caption{Reconstructed $M_{\pi\mu\mu}$ mass distributions of data and MC events passing the $\kpmmlnv$~(a) and $\kpmmlnc$~(b) selections. The signal mass regions are indicated with vertical arrows. Statistical errors on the MC $K_{3\pi}$ component are not indicated and are approximately of the same size as the data errors. The corresponding data/MC ratios with MC statistical errors taken into account are also shown.}\label{fig:mpimm}  
\end{figure}

MC simulations of the $K^{\pm}$ decay channels with three tracks in the final state are used for the background estimation.
The MC events have been generated in a wider range of kaon decay longitudinal coordinate than the fiducial region, to account for event migration due to resolution effects.
The reconstructed $M_{\pi\mu\mu}$ mass distributions of data and MC events passing the $\kpmmlnv$ and $\kpmmlnc$ selections are shown in Fig.~\ref{fig:mpimm}. 
One event is observed in the signal region after applying the $\kpmmlnv$ selection,
while 3489 $\kpimm{\pm}$ candidates are selected with the $\kpmmlnc$ selection.
The expected backgrounds to the $\kpmm$ samples evaluated with MC simulations are reported in Table~\ref{tab:kpimm_bkg}.
For each considered background~$i$, the size of the produced MC sample relative to the expected abundance in data is quantified by the ratio~$\rho_i$:
$$
\rho_i \eqdef \frac{N_{\rm gen}^i}{N_K \mathcal{B}_i},
$$
where $N_{\rm gen}^i$ is the number of MC events generated in the fiducial region and $\mathcal{B}_i$ is the branching fraction of the background~$i$.
\begin{table}[t!]
\begin{center}
\caption{Dominant background contributions to the $\kpmm$ samples:
branching fractions, relative MC sample sizes~$\rho$ and expected numbers of background events~$N_{\rm exp}$ in the $\kpmmlnv$ and $\kpmmlnc$ samples,  obtained from MC simulations. The errors~$\delta N_{\rm exp}$ are dominated by the uncertainties due to limited MC statistics, except for the $\kpimm{\pm}$ background in the $\kpmmlnc$ sample, in which the external error on the branching fraction dominates. 
For the $\kmufour{\pm}$ and $\kmunumumu{\pm}$ decays the ChPT expectation for the branching fractions are used.
The $\mathcal{B}(\kmufour{\pm})$ prediction of Ref.~\cite{kl4-1994} is increased by 8\% to take into account a more precise $\kefour{\pm}$ form factor measurement~\cite{ba10,ba15}. The last row shows the numbers of observed data events for comparison.}
\vspace{0.3cm}
\begin{tabular}{|l|l|l|l|l|}
\hline
Decay channel              & Branching fraction & $\rho$ & $N_{\rm exp}^{\rm LNV} \pm \delta N_{\rm exp}^{\rm LNV}$ & $N_{\rm exp}^{\rm LNC} \pm \delta N_{\rm exp}^{\rm LNC}$ \\
\hline
\hline
$\kthreepic{\pm}$          & $(5.583\pm0.024)  \times  10^{-2}$~\cite{pdg} & $1.16$ & $0.864 \pm 0.864$ & $10.85 \pm 3.06$\\
\hline
$\kmufour{\pm}$            & $(4.5\pm0.2)  \times  10^{-6}$ & $120$  & $0.027 \pm 0.015$ & $0.018 \pm 0.012$\\
                           & (expected~\cite{kl4-1994,ba10,ba15})                     & & & \\
\hline
$\kpimm{\pm}$              & $(9.4\pm0.6)  \times  10^{-8}$~\cite{pdg} & $573$ & $0.257 \pm 0.027$ & $3422 \pm 219$\\
\hline
$\kmunumumu{\pm}$          & $1.35  \times  10^{-8}$  & $3988$ & $0.012 \pm 0.001$ & $0.037 \pm 0.003$\\
                           & (expected~\cite{kl4-1993})                          & & & \\
\hline
Total                      &    $-$                & $-$ & $1.160 \pm 0.865$ & $3433 \pm 219$\\
\hline
\hline
Observed                   &    $-$                & $-$ & $1$ & $3489$\\
\hline

\end{tabular}
\label{tab:kpimm_bkg}
\end{center}
\end{table}
An additional 10\% systematic error due to the limited accuracy of the MC simulation is assigned to the $\kpmmlnv$ total background estimate.
The size of this error is determined from the level of agreement of the data and MC distributions in the $M_{\pi\mu\mu}$ control region $456\mbox{ MeV}/c^2<M_{\pi\mu\mu}<480\mbox{ MeV}/c^2$.

\boldmath
\section{Search for two-body resonances}
\unboldmath
\label{sec:res_search}
A search for two-body resonances in the $\kpmm$ candidates over a range of mass hypotheses is performed
across the distributions of the invariant masses $M_{ij}$ ($ij = \pi\mu, \mu^+\mu^-$). A particle~$X$ produced in $K^{\pm}\to\mu^{\pm}X$ ($K^{\pm}\to\pi^{\pm}X$) decays and decaying promptly to $\pi\mu$ ($\mu^+\mu^-$) would produce a narrow spike in the $M_{\pi\mu}$ ($M_{\mu\mu}$) spectrum. 
MC simulations involving isotropic $X$ decay in its rest frame are used to evaluate the acceptances of the selections~(Sec.~\ref{sec:selection}) for the above decay chains depending on the assumed resonance masses and lifetimes.
The mass step of the resonance scans and the width of the signal mass windows around the assumed mass~$M_X$ are determined by the resolutions $\sigma(M_{ij})$ on the invariant masses $M_{ij}$ ($ij = \pi\mu, \mu^+\mu^-$): the mass step is set to $\sigma(M_{ij})/2$, while the half-width of the signal mass window is $\delta_M(M_{X}) = 2 \sigma(M_{ij})$. Therefore the results obtained in the neighbouring mass hypotheses are correlated.
The dependence of the resolutions $\sigma(M_{ij})$ on the assumed resonance mass~$M_X$ evaluated with MC simulations is approximately $\sigma(M_{ij}) = {0.02 \cdot (M_{X} - M_0^{ij})}$ for the LNC~selection, where $M_0^{ij} = M_i + M_j$ is the mass threshold of the $X\to i j$ decay  ($ij = \pi\mu, \mu^+\mu^-$).
In the LNV selection, the tighter $M_{\pi\mu\mu}$ cut leads to a $15$\% smaller resolution.

\begin{figure}[t]
\begin{center}
\resizebox{0.333\textwidth}{!}{\includegraphics{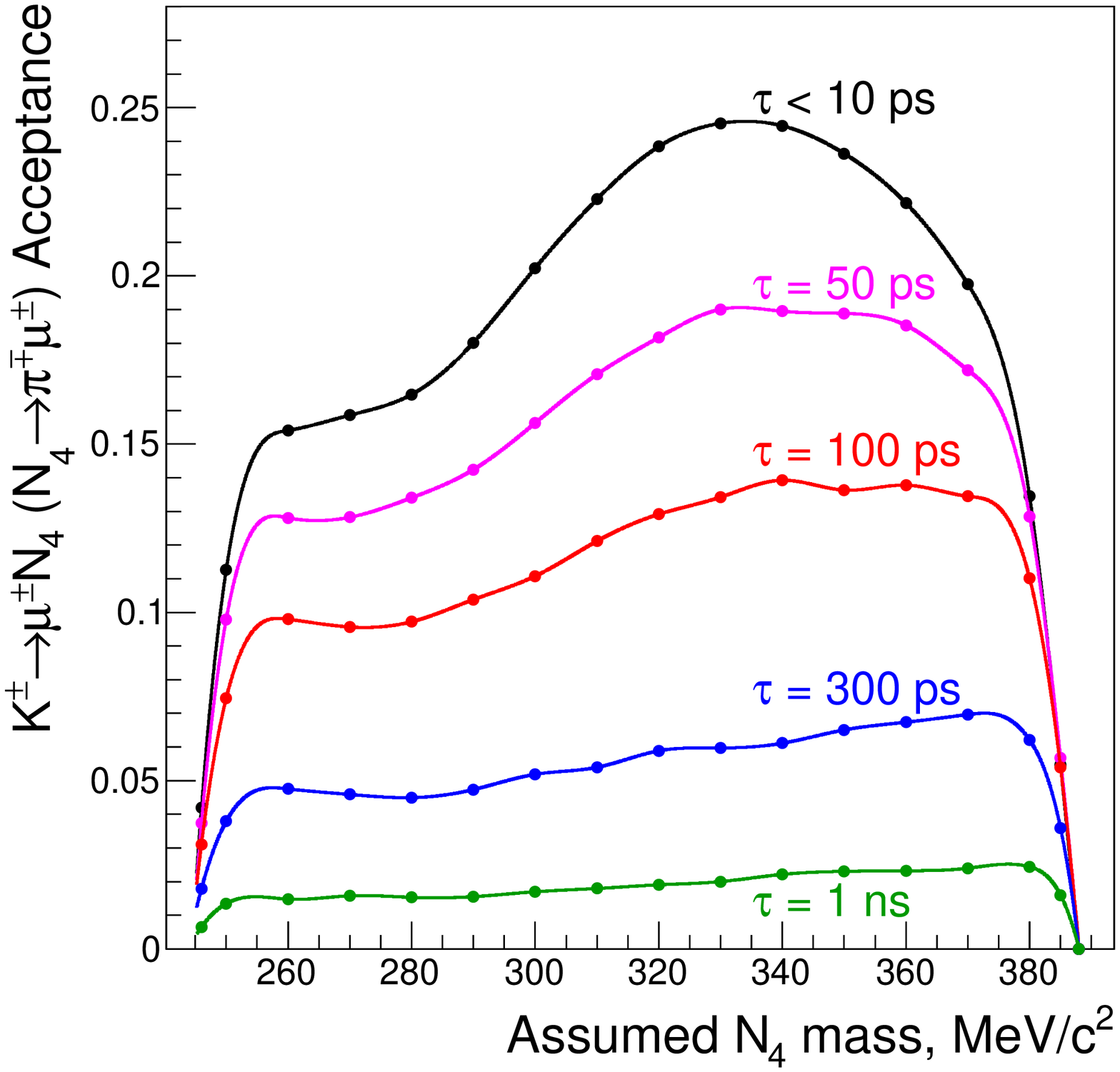}}%
\put(-18,132){(a)}
\resizebox{0.333\textwidth}{!}{\includegraphics{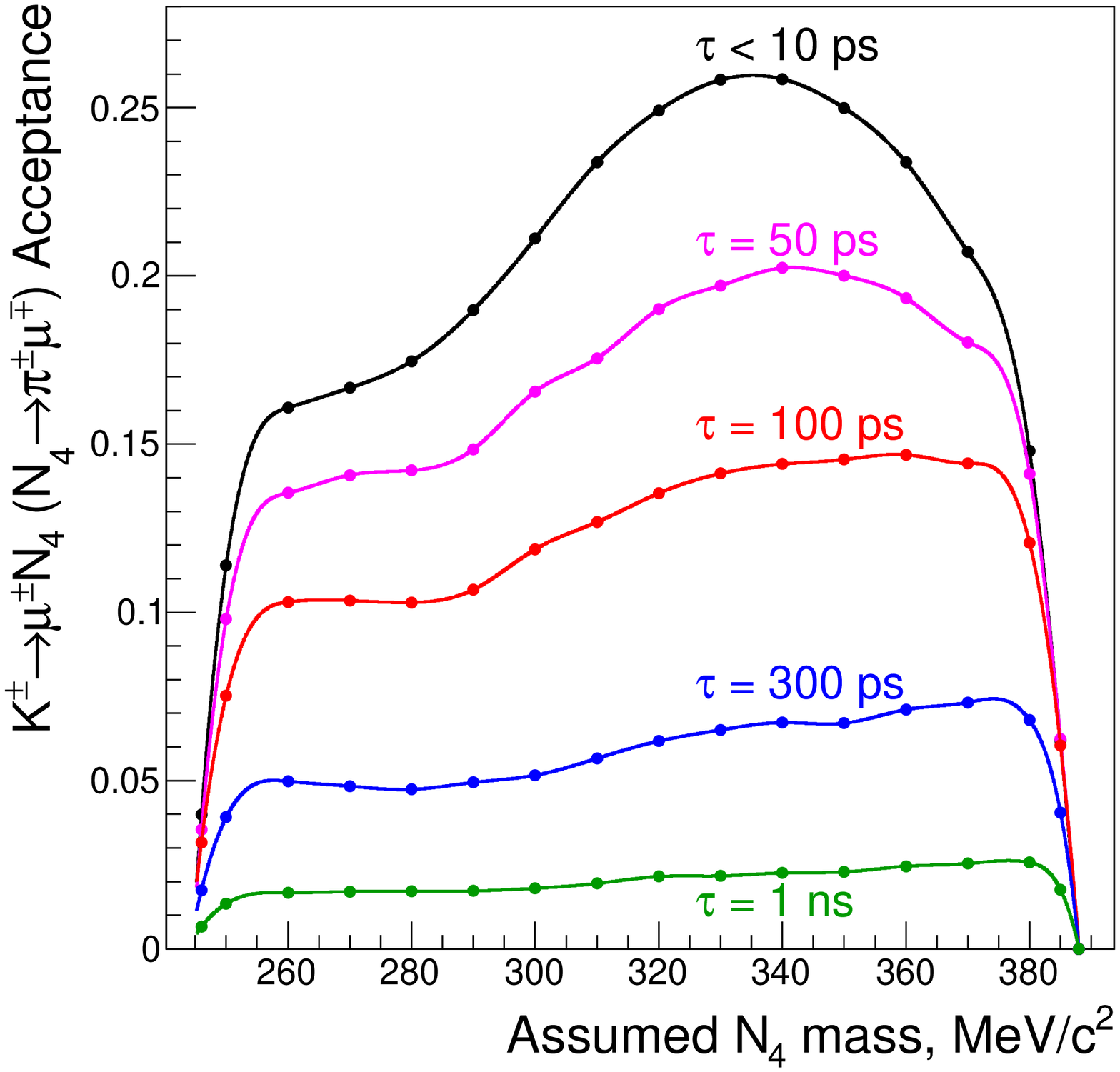}}%
\put(-18,132){(b)}
\resizebox{0.333\textwidth}{!}{\includegraphics{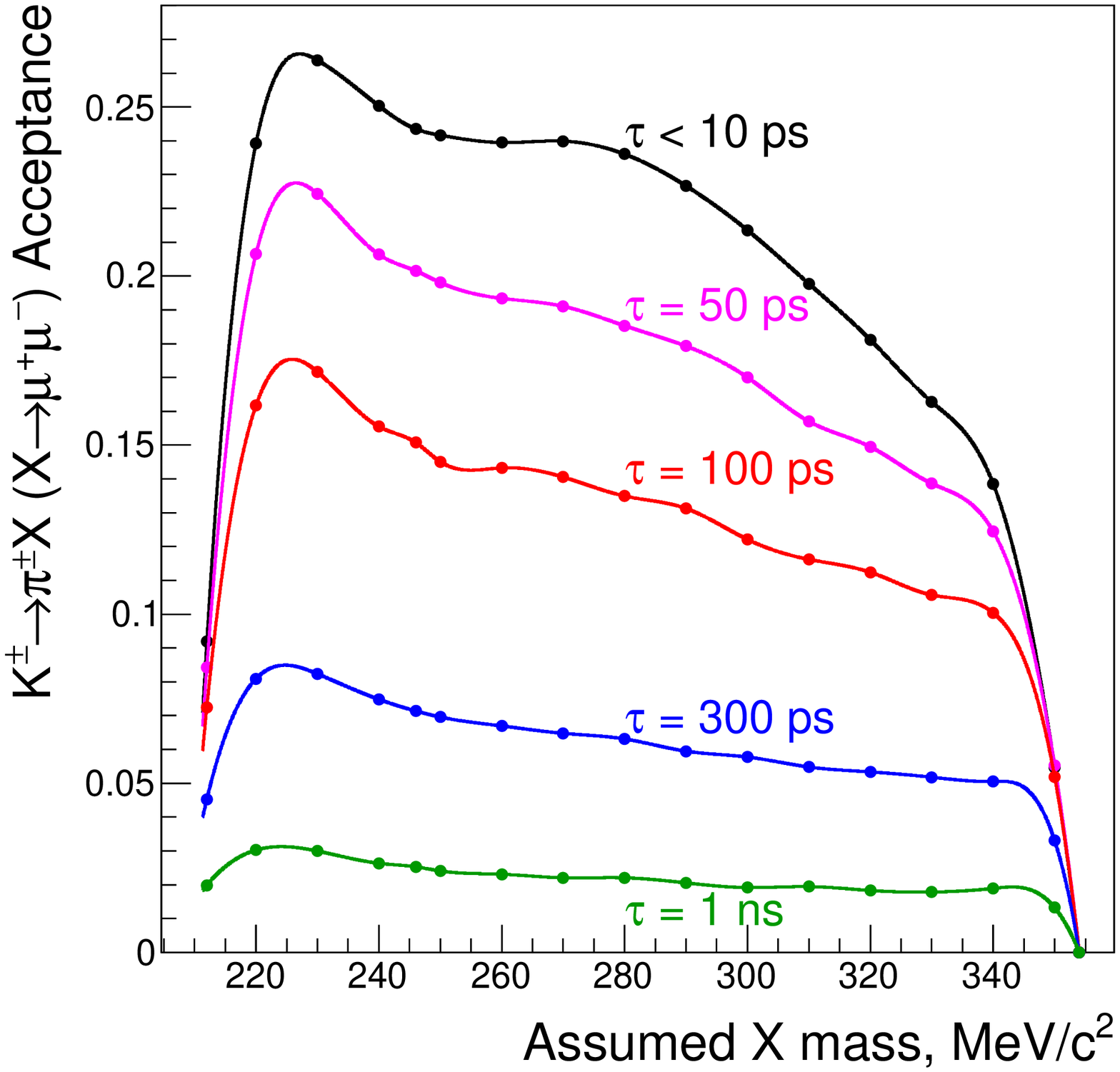}}%
\put(-18,132){(c)}
\end{center}
\vspace{-13mm}
\caption{Acceptances as functions of the assumed resonance mass and lifetime of: (a) the $\kpmmlnv$ selection for $K^{\pm}\to\mu^{\pm}N_4$, $N_4\to\pi^{\mp}\mu^{\pm}$ decays; (b) the $\kpmmlnc$ selection for $K^{\pm}\to\mu^{\pm}N_4$, $N_4\to\pi^{\pm}\mu^{\mp}$ decays; (c) the $\kpmmlnc$ selection for $K^{\pm}\to\pi^{\pm}X$, $X\to\mu^+\mu^-$ decays.
For resonance lifetimes~$\tau>\unit[1]{ns}$ the acceptances scale as $1/\tau$ due to the required three-track vertex topology of the selected events.
In the LNV selection, the tighter $M_{\pi\mu\mu}$ cut leads to a $5$\% smaller acceptance.
The mass dependence in case (c) differs from the others due to the $p>15$~GeV/$c$ pion momentum cut, not applied to muons~(Sec.~\ref{sec:selection}).}\label{fig:acc}  
\end{figure}

The obtained signal acceptances as functions of the resonance mass and lifetime are shown in Fig.~\ref{fig:acc}.
In total, 284 (267) resonance mass hypotheses are tested in the $M_{\pi\mu}$ distribution of the $\kpmmlnv$ ($\kpmmlnc$) candidates and 280 mass hypotheses are tested in the $M_{\mu\mu}$ distribution of the $\kpmmlnc$ candidates, covering the full kinematic range. 

The statistical analysis of the obtained results in each mass hypothesis is performed by applying the Rolke-L\'opez method~\cite{ro01} to find the 90\% confidence intervals for the case of a Poisson process in presence of multiple Poisson backgrounds with unknown mean.
The number of considered backgrounds for the $\kpmmlnv$ ($\kpmmlnc$) candidates is 4 (1); in the latter case, backgrounds other than $\kpimm{\pm}$ are negligible~(Table~\ref{tab:kpimm_bkg}).
Inputs to the Rolke-L\'opez computation are the relative MC sample sizes~$\rho_i$ for each considered background~$i$ and, for each mass hypothesis, 
the number~$N_{\rm obs}$ of observed data events and the number~$N_{\rm bkg}^{\,i}$ of MC events in the signal mass~window.

\boldmath
\section{Results and discussion}
\unboldmath
\label{sec:results}
\boldmath
\subsection{Upper limit on $\mathcal{B}(\kpimmws)$}
\unboldmath
The upper limit~(UL) on the number of $\kpimmws$ signal events in the $\kpmmlnv$ sample corresponding to the observation of one data event and the expected background reported in Table~\ref{tab:kpimm_bkg} is 
$$
 \npmmlnv < 2.92 \quad \mbox{@ 90\% CL}.
$$
Using the value of the signal acceptance~$A(\kpmmlnv)=(20.62\pm0.01)\%$ estimated with MC simulations assuming a uniform phase-space distribution, it leads to an UL on the branching fraction:
$$
\mathcal{B}(\kpimmws) = \frac{\npmmlnv}{N_K\cdot A(\kpmmlnv)}< 8.6 \times 10^{-11} \quad \mbox{@ 90\% CL}.
$$
The total systematic uncertainty on the quoted UL is 1.35\%.
The largest source is the limi\-ted accuracy (Sec.~\ref{sec:datasamples}) of the MC simulations (1.0\%), followed by $\mathcal{B}(\kpimm{\pm})$~(0.75\%),
$\mathcal{B}(\kthreepic{\pm})$~(0.43\%) and $\mathcal{B}(\kmufour{\pm})$~(0.05\%).
The contribution of each component is evaluated as the variation of the result when varying each input separately by one standard deviation.
The contribution of the trigger inefficiency is negligible due to the similar topo\-logy of $\kthreepic{\pm}$ and $\kpimmws$~decays.

\boldmath
\subsection{Limits on two-body resonances}
\begin{figure}[p!]
\begin{center}
\resizebox{0.5\textwidth}{!}{\includegraphics{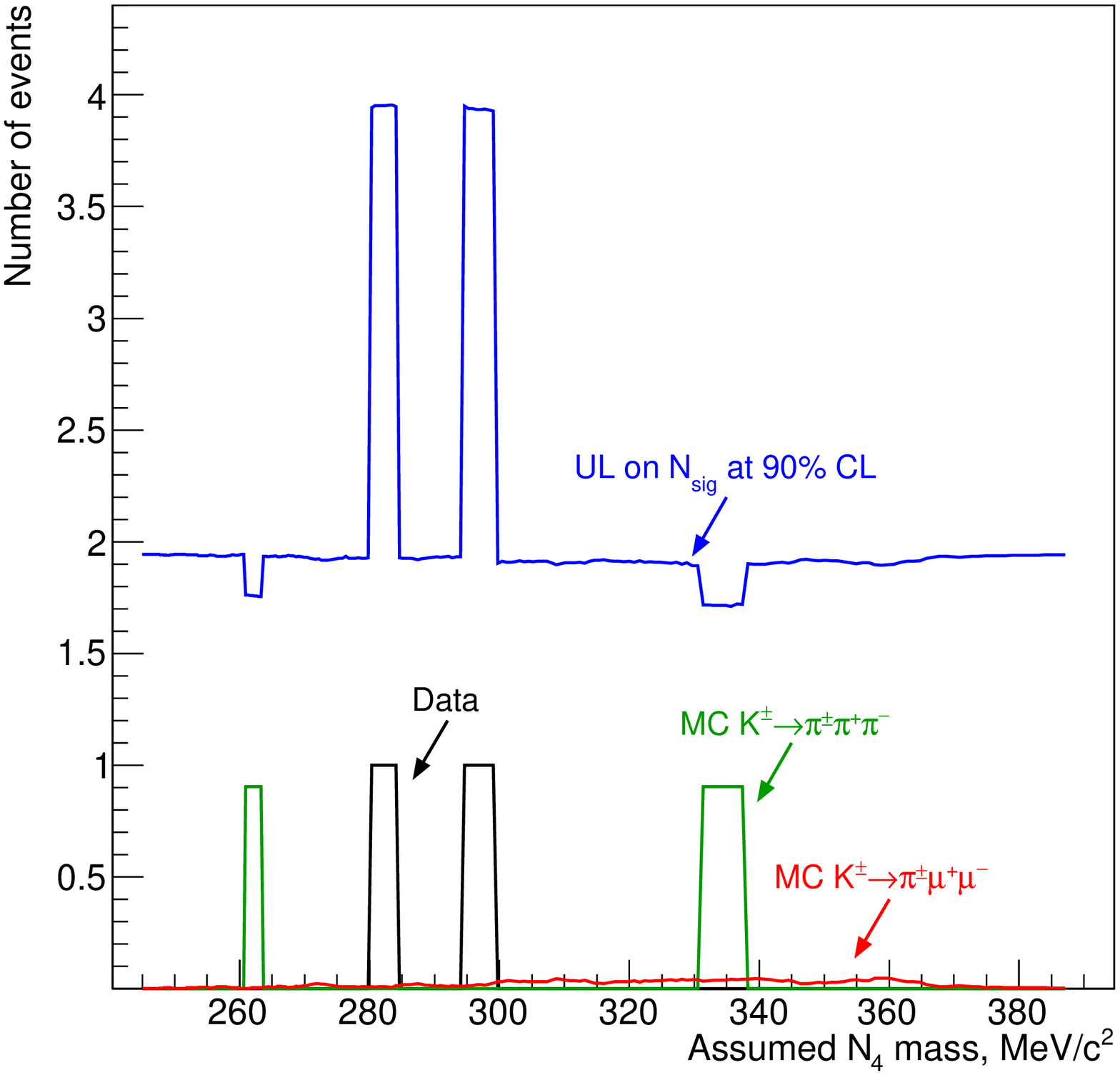}}%
\put(-195,200){(a)}
\resizebox{0.5\textwidth}{!}{\includegraphics{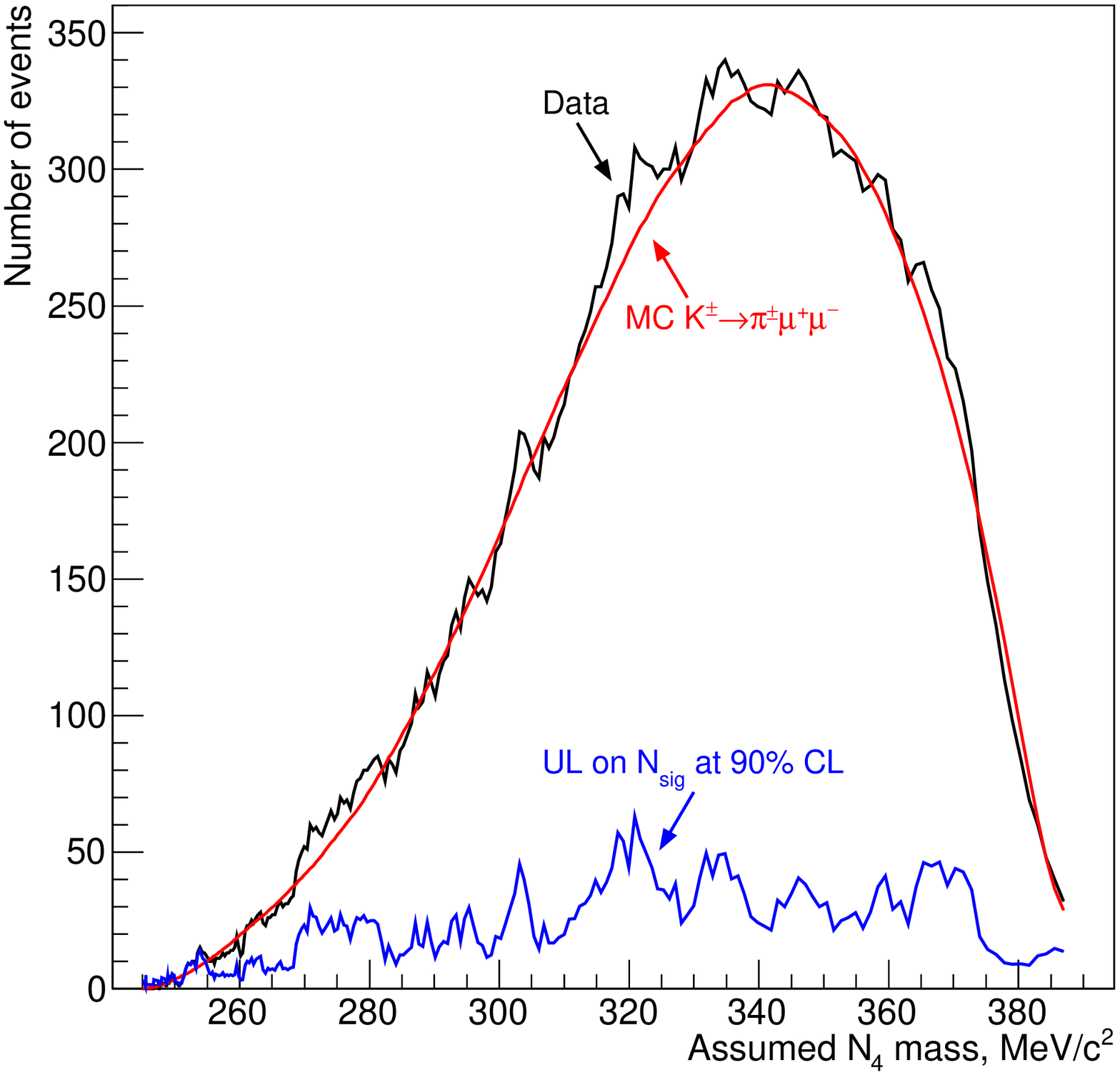}}
\put(-195,200){(b)}\\
\hspace{0.01\textwidth}\resizebox{0.49\textwidth}{!}{\includegraphics{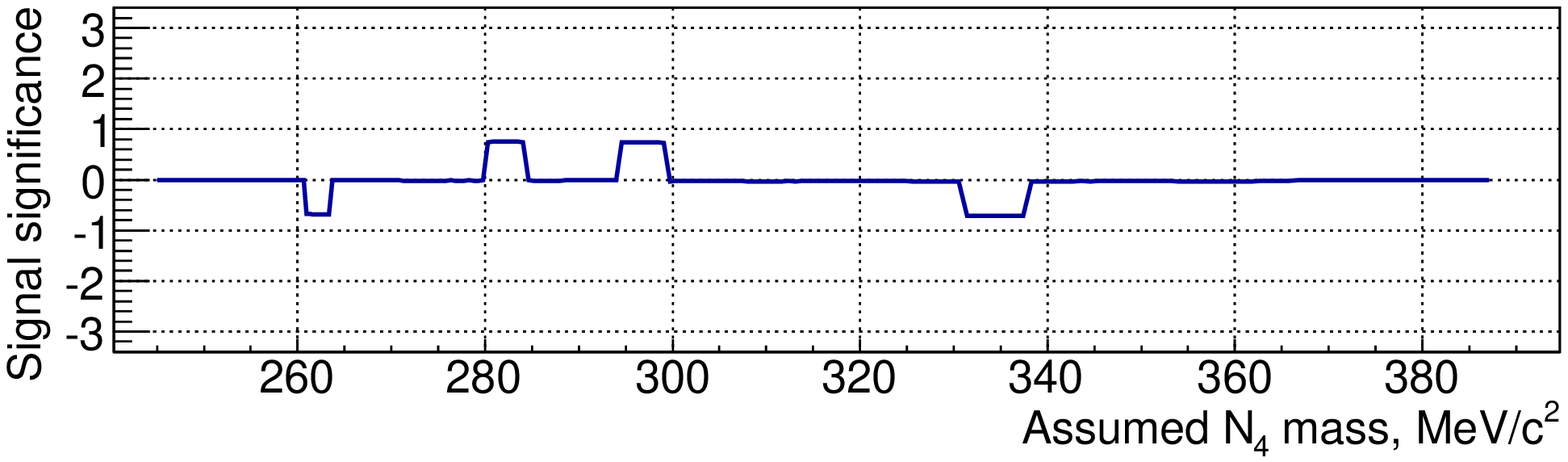}}%
\hspace{0.01\textwidth}\resizebox{0.49\textwidth}{!}{\includegraphics{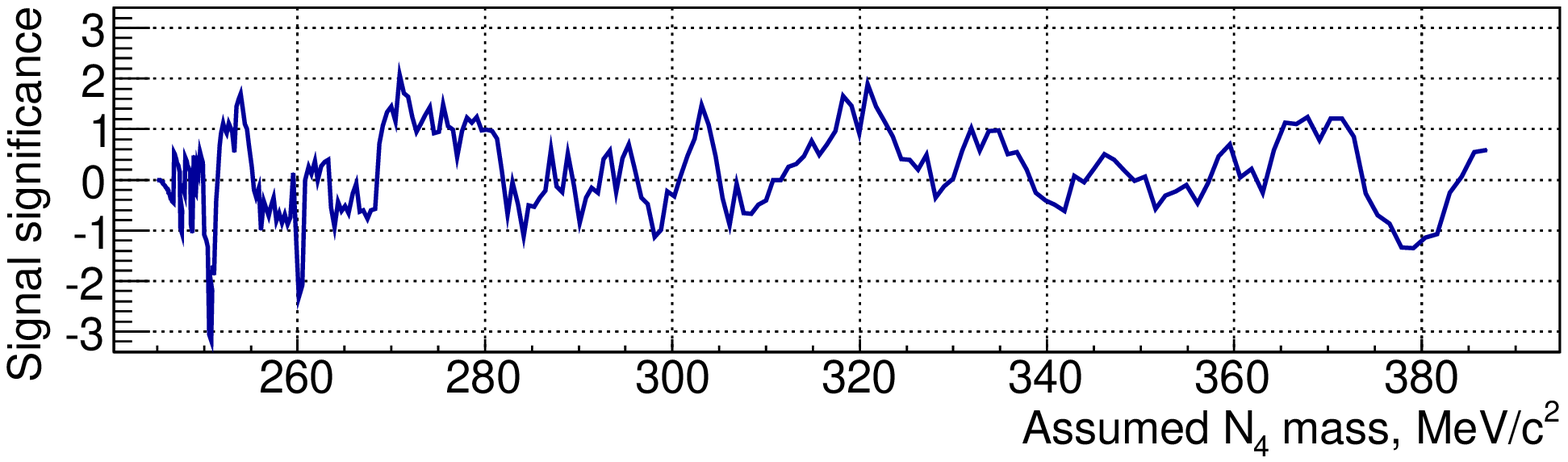}}\\
\resizebox{0.5\textwidth}{!}{\includegraphics{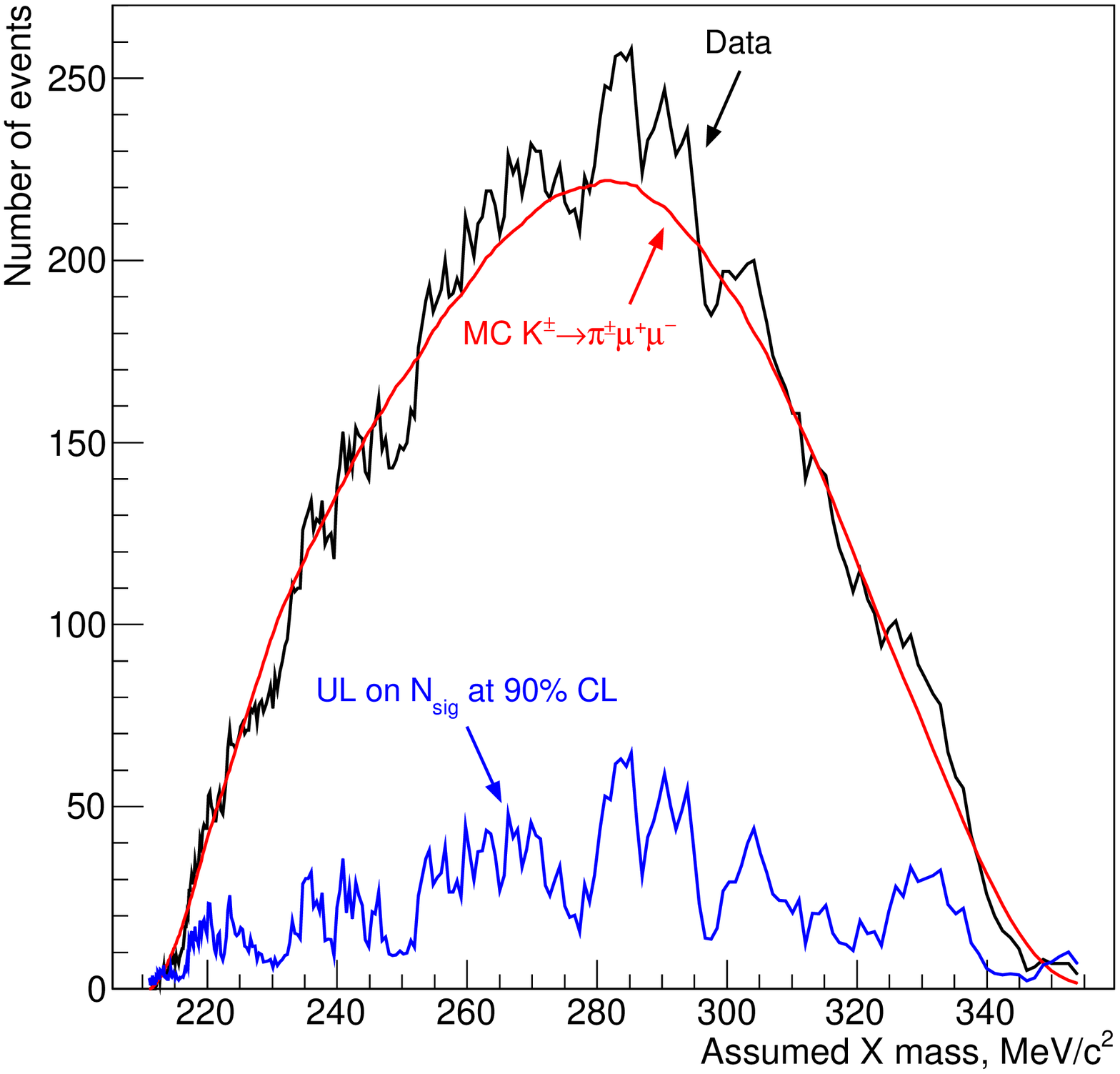}}
\put(-195,200){(c)}\\
\hspace{0.01\textwidth}\resizebox{0.49\textwidth}{!}{\includegraphics{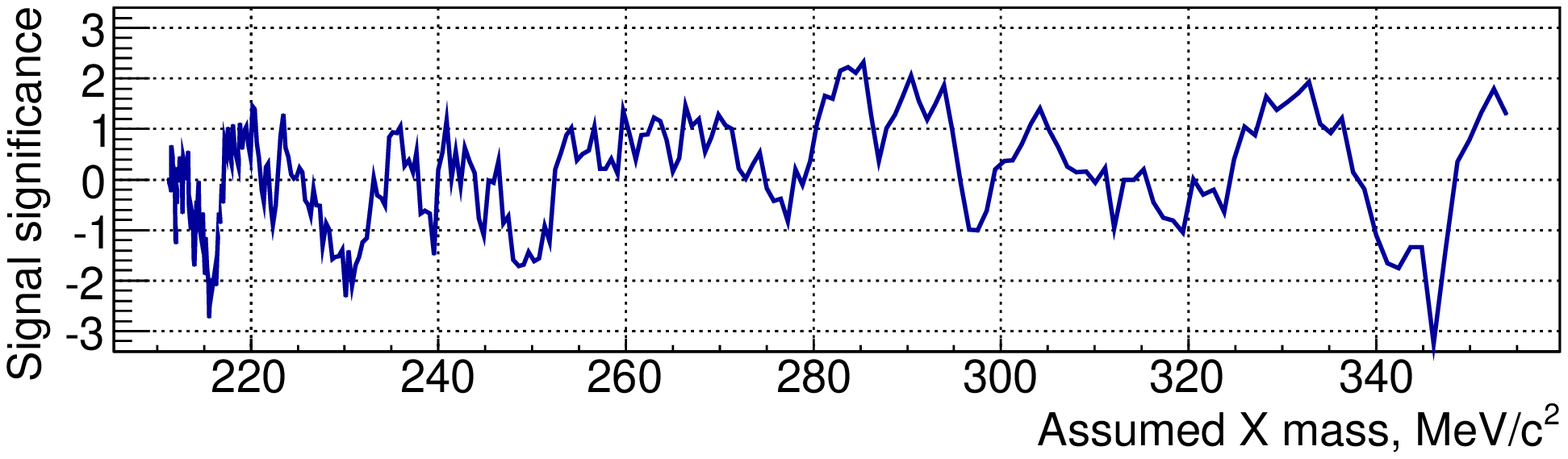}}%
\end{center}
\vspace{-13mm}
\caption{Numbers~$N_{\rm obs}$ of observed data events (Data) and expected background events (MC $\kthreepic{\pm}$ and MC $\kpimm{\pm}$) passing: (a) the $M_{\pi\mu}$ cut with the $\kpmmlnv$ selection; (b) the $M_{\pi\mu}$ cut with the $\kpmmlnc$ selection; (c) the $M_{\mu\mu}$ cut with the $\kpmmlnc$ selection.
The obtained ULs at 90\% CL on the numbers of signal events~$N_{\rm sig}$ and the local significances~$z$ of the signal are also shown for each resonance mass hypothesis. All presented quantities are correlated for neighbouring resonance masses as the mass step of the scans is about 8 times smaller than the signal window width.}\label{fig:nres_ul}  
\end{figure}

\unboldmath
For each of the three resonance searches performed, the local significance~$z$ of the signal is evaluated for each mass hypothesis as
$$
z = \frac{N_{\rm obs}-N_{\rm exp}}{\sqrt{\delta N^2_{\rm obs}+\delta N^2_{\rm exp}}},
$$
where $N_{\rm obs}$ is the number of observed events, $N_{\rm exp}$ is the number of expected background events, $\delta {N_{\rm obs}} = \sqrt{N_{\rm obs}}$, and 
$\delta {N_{\rm exp}} = \sqrt{\sum_i (N_{\rm bkg}^i/\rho_i^2)}$ is the statistical uncertainty on $N_{\rm exp}$ due to the limited size of the MC samples. 
In case $N_{\rm obs}$~($N_{\rm bkg}^i)=0$,
$N_{\rm obs}$~($N_{\rm bkg}^i)=1$ is used for the computation of $\delta {N_{\rm obs}}$~($\delta {N_{\rm exp}}$).
The values~$N_{\rm obs}$, the normalized numbers of background events~$N_{\rm bkg}^{\,i}/\rho_i$, the ULs at 90\% CL on the numbers of signal events and the corresponding local significances~$z$ of the signals are shown for each mass hypothesis in Fig.~\ref{fig:nres_ul}. The local significances never exceed 3 standard deviations: no signal observation is reported.

\begin{figure}[p!]
\begin{center}
\resizebox{0.5\textwidth}{!}{\includegraphics{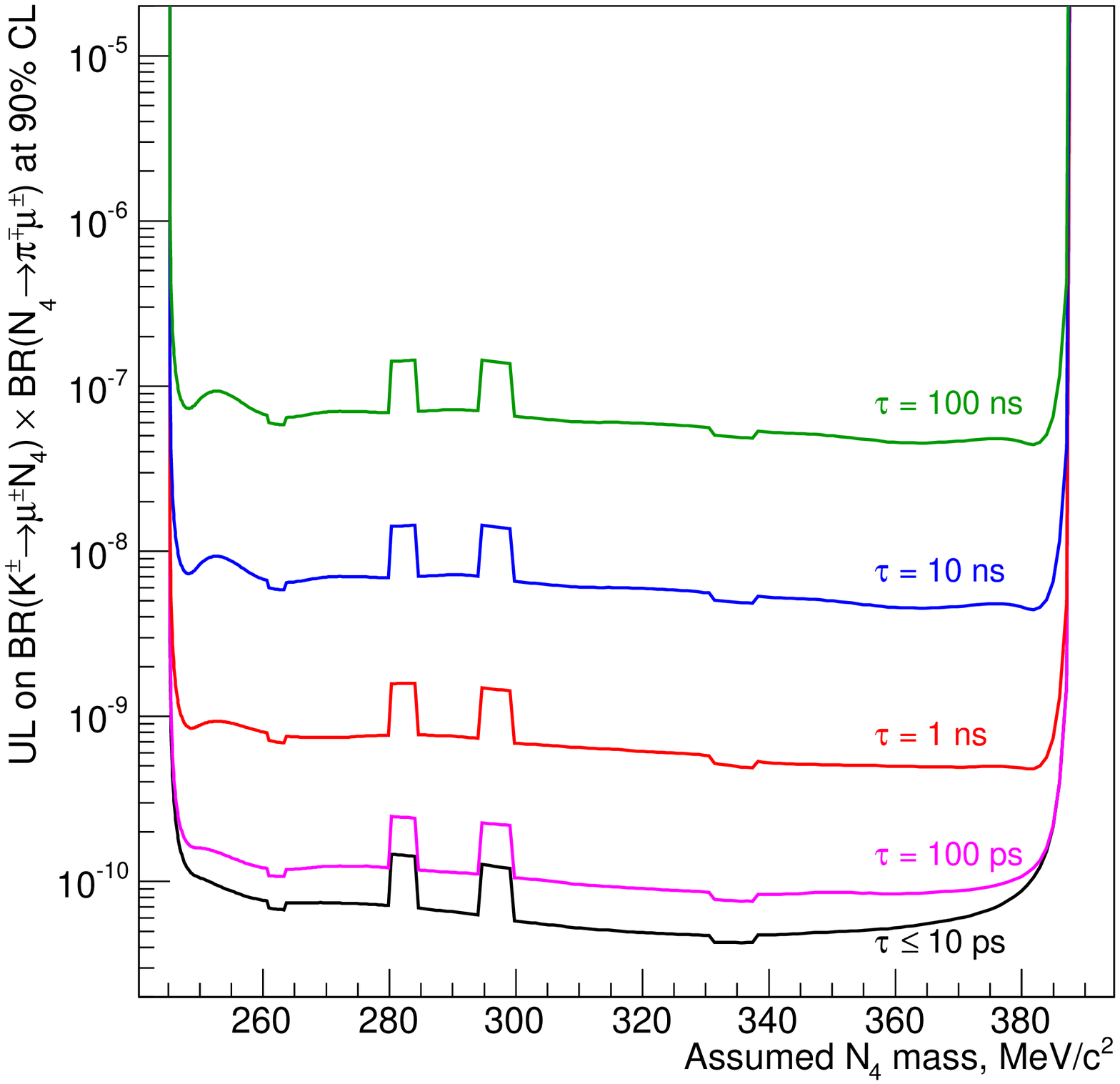}}%
\put(-188,200){(a)}
\resizebox{0.5\textwidth}{!}{\includegraphics{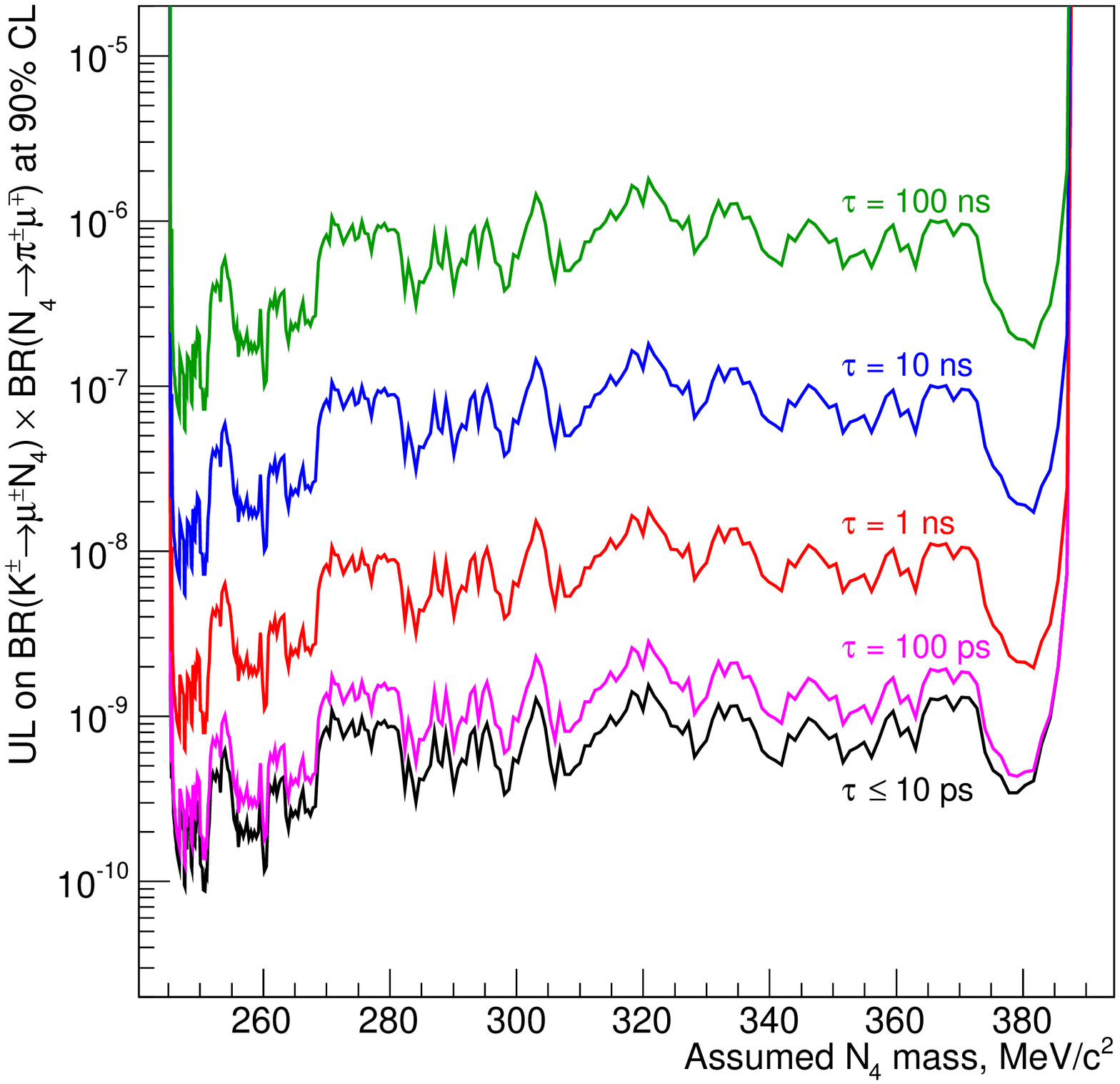}}%
\put(-188,200){(b)}\\
\resizebox{0.5\textwidth}{!}{\includegraphics{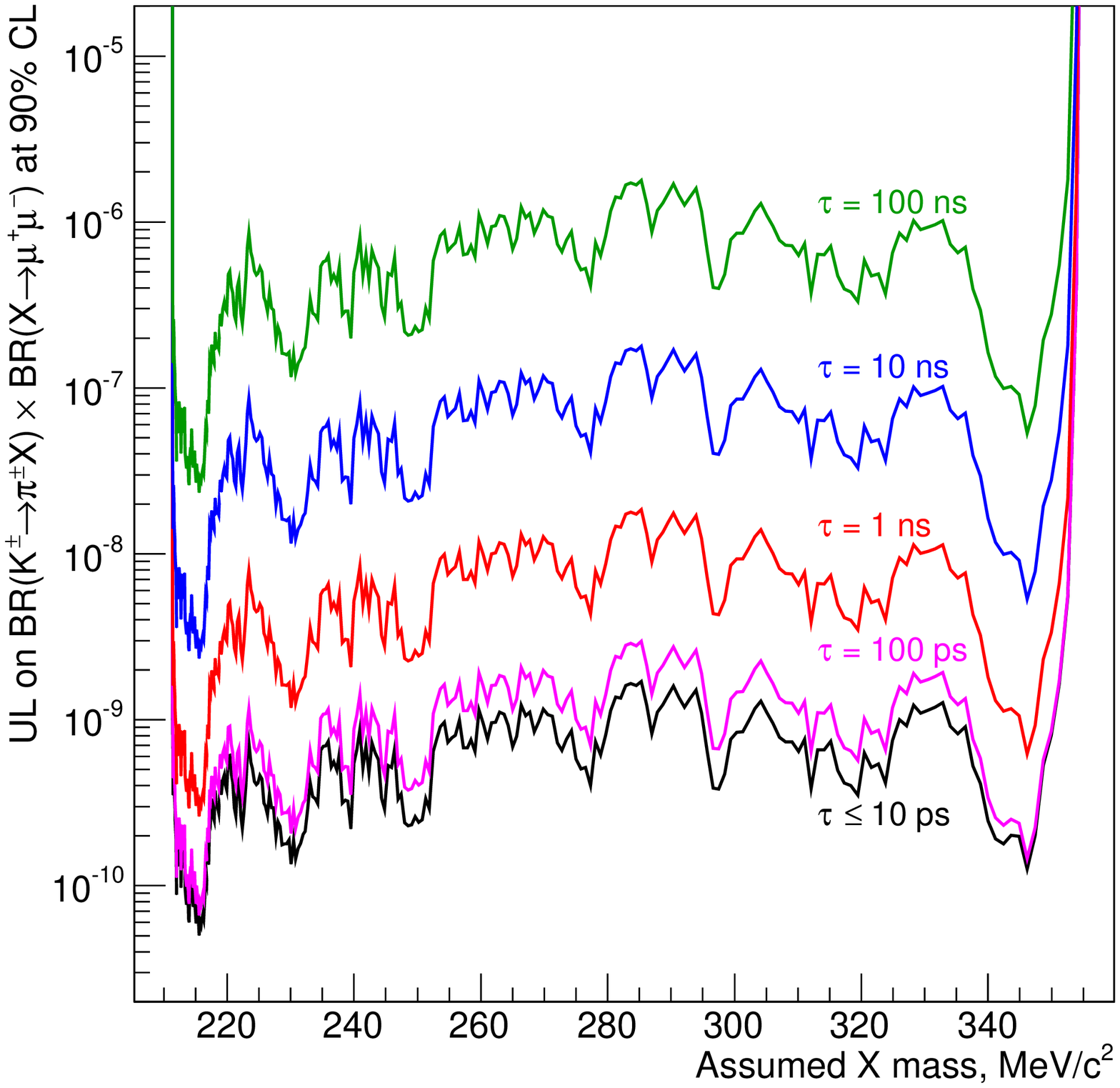}}%
\put(-188,200){(c)}
\end{center}
\vspace{-13mm}
\caption{Obtained ULs at 90\% CL on the products of branching fractions as functions of the assumed resonance mass and lifetime: (a)~$\mathcal{B}(\kmutwoN{\pm})\mathcal{B}(\Npimuws)$; (b)~$\mathcal{B}(\kmutwoN{\pm})\mathcal{B}(\Npimurs)$; (c)~$\mathcal{B}(\kpiX{\pm})\mathcal{B}(\Xmumu)$.}\label{fig:kpimmws_results_data}  
\end{figure}

The ULs on the product~$\mathcal{B}(K^{\pm}\to p_1 X)\mathcal{B}(X\to p_2 p_3)$, $p_1p_2p_3 = \mu^{\pm}\pi^{\mp}\mu^{\pm}, \mu^{\pm}\pi^{\pm}\mu^{\mp},\pi^{\pm}\mu^{+}\mu^{-}$, as functions of the resonance lifetime~$\tau$ are obtained for each mass hypothesis~$M_i$ using the values of the acceptances~$A_{\pi\mu\mu}(M_i,\tau)$ (Fig.~\ref{fig:acc}) and the ULs on the numbers of signal events~$N^i_{\rm sig}$ for that mass hypothesis~(Fig.~\ref{fig:nres_ul}):
$$
\left.\mathcal{B}(K^{\pm}\to p_1 X)\mathcal{B}(X\to p_2 p_3)\right|_{M_i,\tau} = \frac{N_{\rm sig}^i}{N_K\cdot A_{\pi\mu\mu}(M_i,\tau)}.
$$
The obtained ULs as functions of the resonance mass, for several values of the assumed resonance lifetime, are shown in Fig.~\ref{fig:kpimmws_results_data}.
The largest source of systematic uncertainty on the ULs for lifetimes $\tau \leq 10$~ns
is the limited precision of $N_K$ (0.4\%), while for $\tau = 100$~ns the uncertainty (3\%) due to the limited size of the MC sample used for the acceptance evaluation dominates.
The systematic uncertainties on $N_{\rm sig}^i$ are negligible: for the $\kpmmlnv$ sample, the expected background is negligible in most of the mass hypotheses; for the $\kpmmlnc$ sample, the $\kpimm{\pm}$ MC simulation is scaled to match the data, such that it does not rely on the measurements of $\mathcal{B}(\kpimm{\pm})$ and the form factor~\cite{ba11}, which were obtained with a sample of comparable size of the present one.
Other systematic errors (e.g. residual background contamination) are negligible.

\boldmath
\subsection{Interpretation of $\mathcal{B}(\kmutwoN{\pm})\mathcal{B}(\Npimuns)$}
\begin{figure}[h]
\begin{center}
\resizebox{0.5\textwidth}{!}{\includegraphics{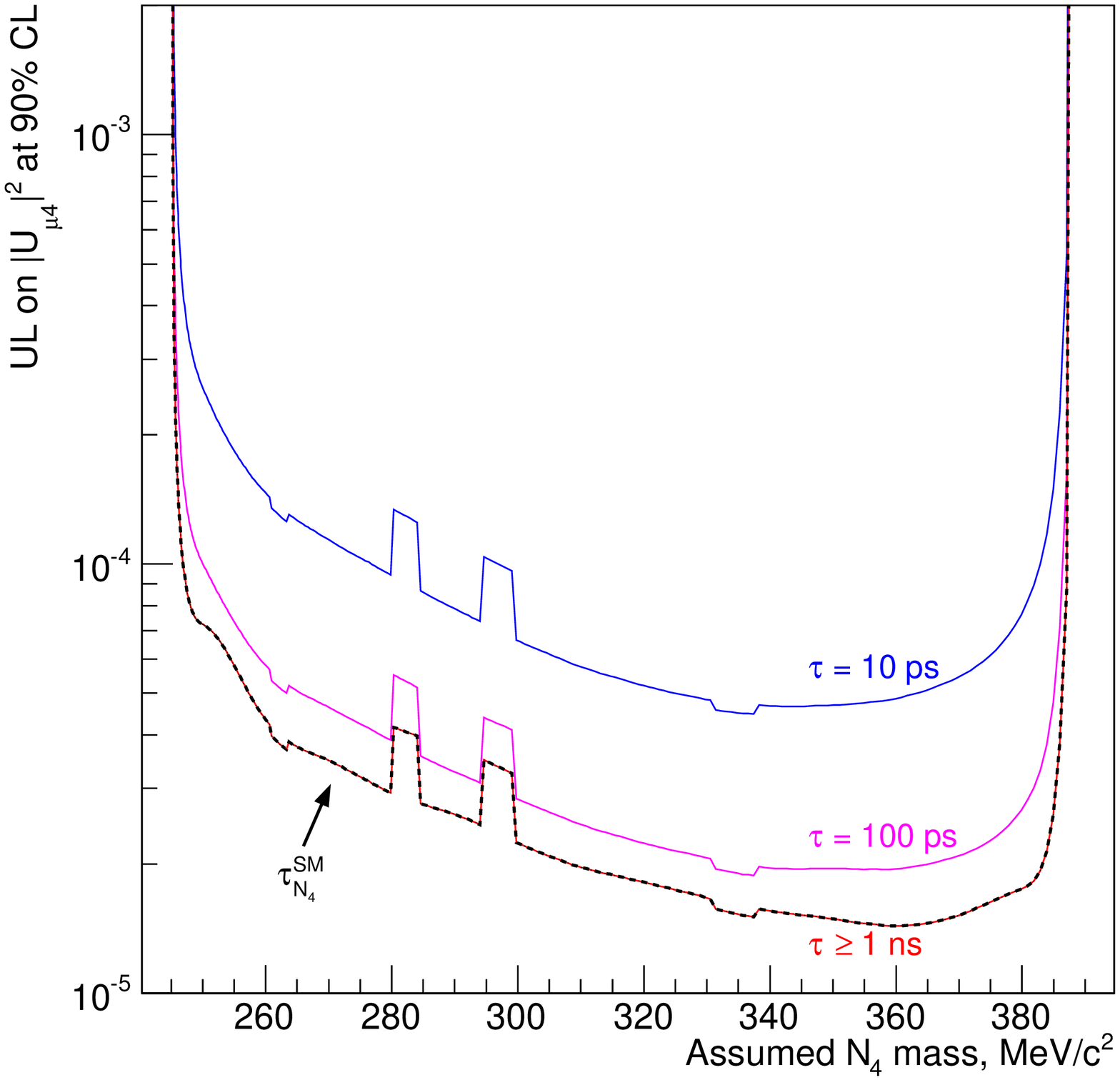}}%
\put(-188,200){(a)}
\resizebox{0.5\textwidth}{!}{\includegraphics{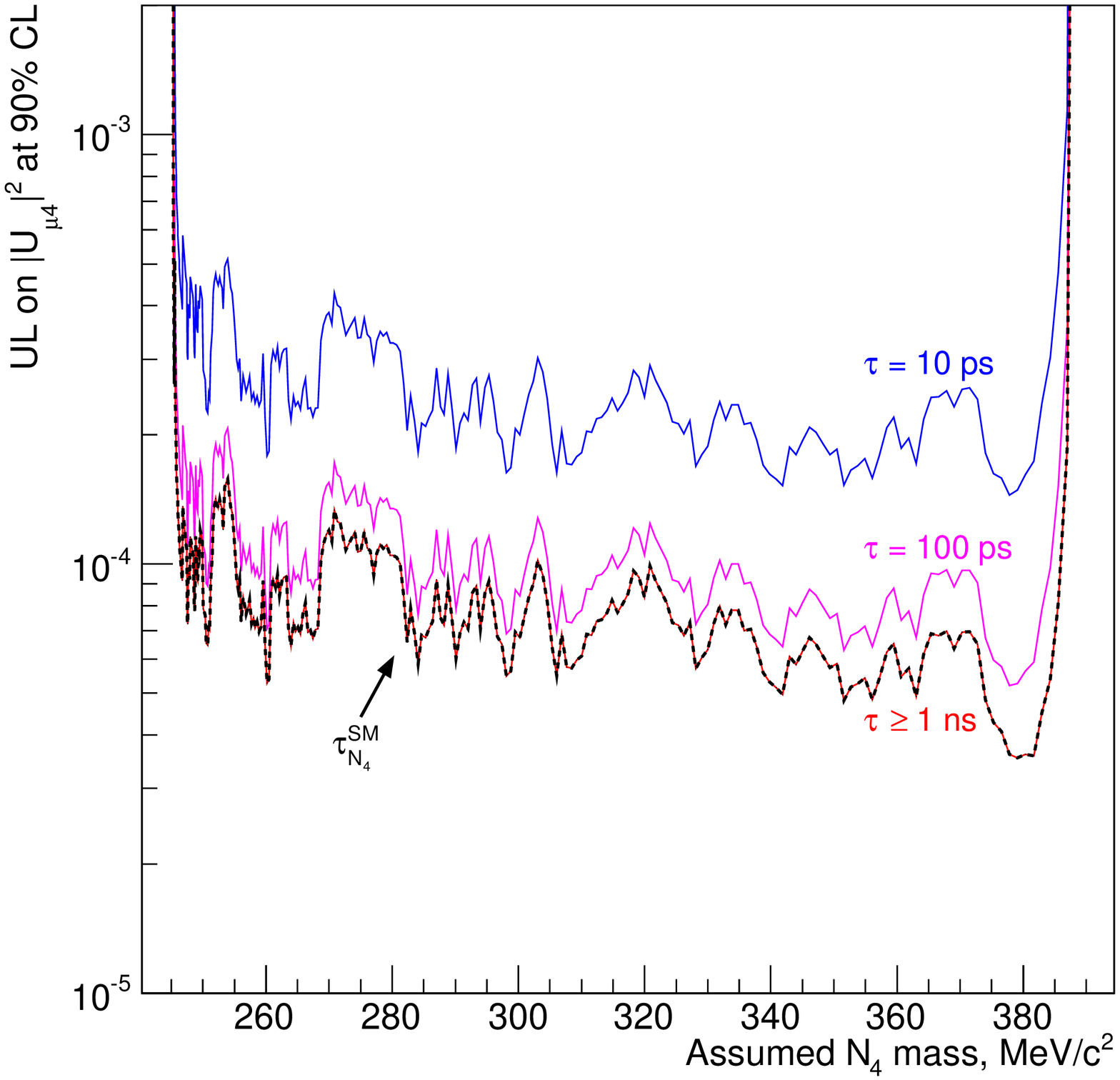}}%
\put(-188,200){(b)}
\end{center}
\vspace{-13mm}
\caption{Upper limits at 90\% CL on $|U_{\mu 4}|^2$ as functions of the assumed resonance mass and lifetime obtained from the limits on: (a)~$\mathcal{B}(\kmutwoN{\pm})\mathcal{B}(\Npimuws)$; (b)~${\mathcal{B}(\kmutwoN{\pm})\mathcal{B}(\Npimurs)}$. The boundaries for $\tau \geq 1$~ns are valid up to a maximum lifetime of $\sim 100$~$\mu$s.}\label{fig:coupling_neutrinos}  
\end{figure}

\unboldmath
Limits on the products~$\mathcal{B}(\kmutwoN{\pm})\mathcal{B}(\Npimuns)$ obtained from $\kpmmlnv$ and $\kpmmlnc$ samples can be
used to constrain the squared magnitude $|U_{\mu4}|^2$ using the relation~\cite{cv10}
$$
|U_{\mu4}|^2 = \frac{8\sqrt{2}\pi\hbar}{G_F^2 \sqrt{M_K \tau_K} f_Kf_{\pi}|V_{us}V_{ud}|} \sqrt{\frac{\mathcal{B}(\kmutwoN{\pm})\mathcal{B}(\Npimuns)}{\tau_{N_4} M_{N_4}^5\lambda^{\frac{1}{2}}(1,r_{\mu}^2,r_{N_4}^2)\lambda^{\frac{1}{2}}\left(1,\rho_{\pi}^2,\rho_{\mu}^2\right)\chi_{\mu\mu}}},
$$
where $r_{i} \eqdef M_{i}/M_K$, $\rho_i \eqdef M_{i}/M_{N_4}$ ($i=\mu,\pi, N_4$), $\lambda(a,b,c) \eqdef a^2 + b^2 + c^2 -2ab -2ac -2bc$ and $\chi_{\mu\mu} = [(1+\rho_{\mu}^2)-(r_{N_4}^2-r_{\mu}^2)(1-\rho_{\mu}^2)][(1-\rho_{\mu}^2)^2-(1+\rho_{\mu}^2)\rho_{\pi}^2]$.
The value of the lifetime~$\tau_{N_4}^{\ttiny{SM}}$, obtained assuming that the heavy neutrino decays into SM particles only and that $|U_{e4}|^2 = |U_{\mu4}|^2 = |U_{\tau4}|^2$, is evaluated for each mass hypothesis, using the decay widths provided in Ref.~\cite{at09}.
The ULs on $|U_{\mu4}|^2$ as functions of the resonance mass obtained for several values of the assumed resonance lifetime, including~$\tau_{N_4}^{\ttiny{SM}}$, are shown in Fig.~\ref{fig:coupling_neutrinos}.

\boldmath
\subsection{Interpretation of $\mathcal{B}(\kpiX{\pm})\mathcal{B}(\Xmumu)$}
\begin{figure}[h]
\begin{center}
\resizebox{0.5\textwidth}{!}{\includegraphics{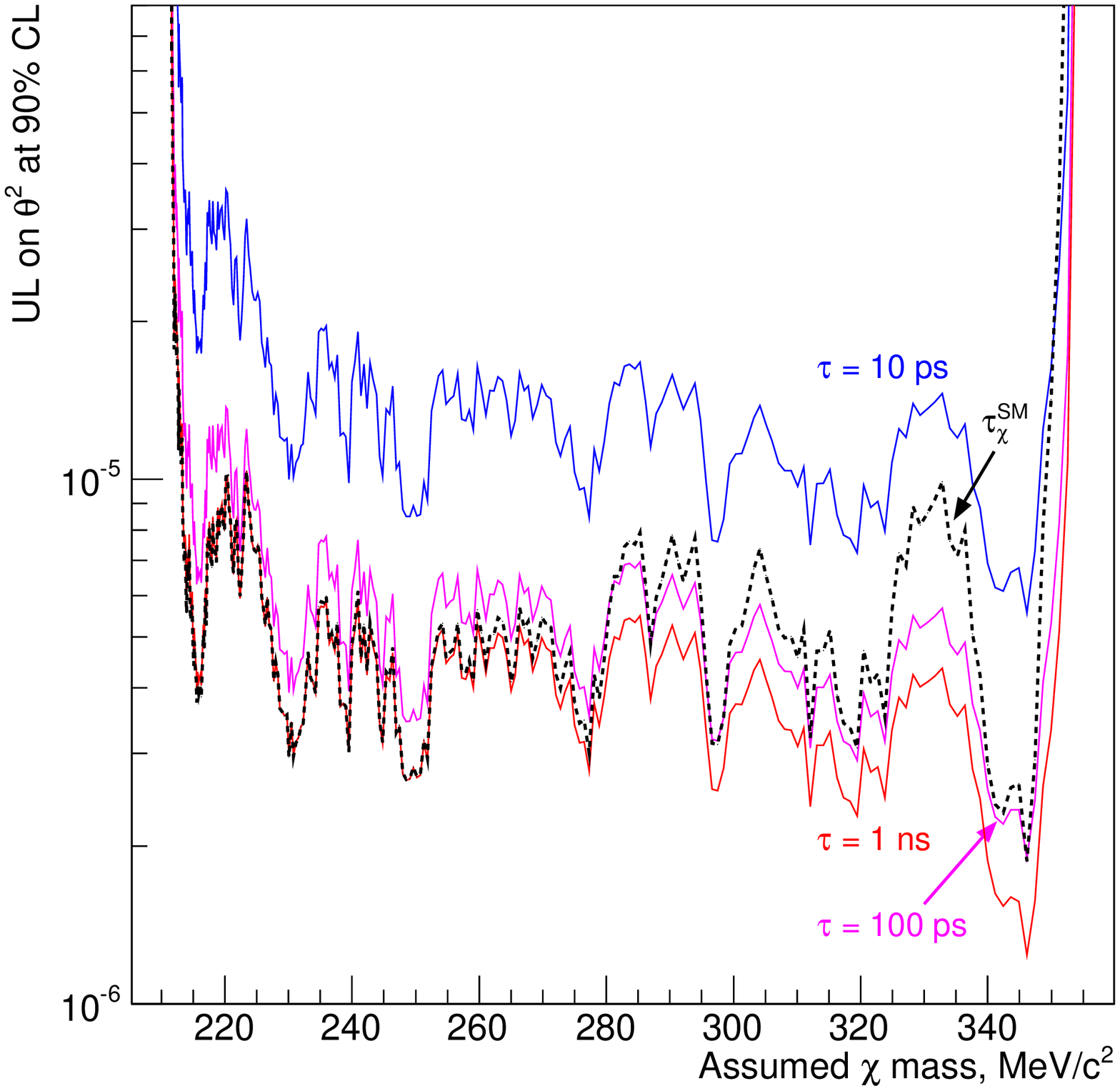}}%
\end{center}
\vspace{-13mm}
\caption{Upper limits at 90\% CL on the squared Higgs-inflaton mixing angle~$\theta^2$ as functions of the resonance mass and lifetime obtained from the limits on~$\mathcal{B}(\kpiX{\pm})\mathcal{B}(\Xmumu)$. The UL corresponding to the lifetime $\tau_{\chi}^{\ttiny{SM}}$ moves across the ones corresponding to fixed lifetimes as $\tau_{\chi}^{\ttiny{SM}}$ becomes smaller for larger inflaton masses.}\label{fig:coupling_inflatons}  
\end{figure}

\unboldmath
The obtained UL on the product~$\mathcal{B}(\kpiX{\pm})\mathcal{B}(\Xmumu)$ can be
used to constrain the magnitude of the inflaton-Higgs mixing angle~$\theta$ using the relation~\cite{be10}
$$
\theta^2 = \sqrt{\frac{8\pi \hbar v^2}{\alpha_{\chi}}}\sqrt{\frac{\mathcal{B}(\kpichi{\pm})\mathcal{B}(\chimumu)}{\tau_{\chi}M_{\chi}^3 \lambda^{\frac{1}{2}}(1,r_{\pi}^2,r_{\chi}^2)\lambda^{\frac{1}{2}}\left(1,\tilde\rho_{\mu}^2,\tilde\rho_{\mu}^2\right)\tilde\chi_{\mu\mu}}},
$$
where $\tilde \rho_i \eqdef M_{i}/M_{\chi}$ ($i=\mu,\pi$), $\alpha_{\chi} \approx 1.3\times 10^{-3}$ and $\tilde \chi_{\mu\mu} = \tilde\rho_{\mu}^2(1-4\tilde\rho_{\mu}^2)$.
The value of the lifetime~$\tau_{\chi}^{\ttiny{SM}}$, obtained assuming that the inflaton decays into SM particles only, is evaluated for each mass hypothesis, using the decay widths provided in Ref.~\cite{be10}.
The ULs on $\theta^2$ as functions of the resonance mass obtained for several values of the assumed resonance lifetime, including~$\tau_{\chi}^{\ttiny{SM}}$, are shown in Fig.~\ref{fig:coupling_inflatons}.
 
\boldmath
\section*{Conclusions}
\unboldmath
Searches for the LNV $\kpimmws$ decay and resonances in $\kpimmns{\pm}$ decays at the NA48/2 experiment with the 2003--2004 data are presented. 
No signals are observed. An UL of $8.6\times10^{-11}$ on $\mathcal{B}(\kpimmws)$ is established at 90\% CL, improving the previous limit~\cite{ba11} by more than one order of magnitude.

Upper limits are set on the products of branching fractions~$\mathcal{B}(\kmutwoN{\pm})\mathcal{B}(\Npimuns)$ and $\mathcal{B}(\kpiX{\pm})\mathcal{B}(\Xmumu)$ as functions of the assumed resonance mass and lifetime. These limits are in the $(10^{-11},10^{-9})$ range for resonance lifetimes below 100~ps.
Using these constraints, ULs on heavy neutrino and inflaton parameters~$|U_{\mu 4}|^2$ and $\theta^2$ are obtained as functions of the resonance mass and lifetime. 


\boldmath
\section*{Acknowledgements}
\unboldmath

We express our gratitude to the staff of the CERN laboratory and the technical staff of the participating universities and laboratories for their efforts in the operation of the experiment and data processing. We thank Dmitry Gorbunov for useful discussions.


\end{document}